\documentclass[11pt]{article}
\usepackage{graphicx,amsmath,amsfonts,amssymb,dcolumn,amsthm}

\begin{document}

\title{\textbf{Gribov ambiguities at the Landau - maximal Abelian interpolating gauge}}
\author{\textbf{Ant\^onio D.~Pereira Jr.}\thanks{aduarte@if.uff.br} \  and \  \textbf{Rodrigo F.~Sobreiro}\thanks{sobreiro@if.uff.br}\\\\
\textit{{\small UFF $-$ Universidade Federal Fluminense,}}\\
\textit{{\small Instituto de F\'{\i}sica, Campus da Praia Vermelha,}}\\
\textit{{\small Avenida General Milton Tavares de Souza s/n, 24210-346,}}\\
\textit{{\small Niter\'oi, RJ, Brasil.}}}
\date{}
\maketitle

\begin{abstract}
\noindent In a previous work, we presented a new method to account for the Gribov ambiguities in non-Abelian gauge theories. The method consists on the introduction of an extra constraint which directly eliminates the infinitesimal Gribov copies without the usual geometric approach. Such strategy allows to treat gauges with non-hermitian Faddeev-Popov operator. In this work, we apply this method to a gauge which interpolates among the Landau and maximal Abelian gauges. The result is a local and power counting renormalizable action, free of infinitesimal Gribov copies. Moreover, the interpolating tree-level gluon propagator is derived.
\end{abstract}


\section{Introduction}\label{INTRO}

One of the most important and challenging open problems in theoretical Physics is the full comprehension of the non-perturbative features of Yang-Mills theories. Responsible for describing the successful Standard Model at high energies, Yang-Mills theories still lack a complete consistent quantization. As pointed out by V.~N.~Gribov \cite{Gribov:1977wm} at the Landau gauge, a residual gauge symmetry survives the Faddeev-Popov gauge fixing procedure \cite{Faddeev:1967fc}. It is a known fact that, to quantize a gauge theory, it is necessary to consistently eliminate the gauge freedom of the Yang-Mills action, see also \cite{Sobreiro:2005ec}. The residual gauge symmetry is characterized by the presence of redundant configurations (called Gribov copies) which still contribute to the path integral. A very important remark is that it is not a particular defect of Landau gauge, but of all covariant gauges, as formally shown by I.~M.~Singer \cite{Singer:1978dk}. Since these configurations represent a redundancy in the theory, their elimination is an unavoidable requirement.

Still in \cite{Gribov:1977wm}, Gribov showed that copies which are related by infinitesimal gauge transformations are associated with the zero-modes of the Faddeev-Popov operator (or, equivalently, to the poles of the ghost propagator) for the Landau gauge. In fact, this is true at least for all gauges that depend exclusively on the gauge field, see \cite{Pereira:2013aza}. Moreover, Gribov proposed the elimination of these infinitesimal copies by restricting the path integral to a region which is free of infinitesimal copies. This region is known as the first Gribov region or, simply, Gribov region. Essentially, it is defined as the region where the Faddeev-Popov operator is positive-definite, a property that ensures that no infinitesimal copies are present. A very important feature is that all gauge orbits actually cross the Gribov region \cite{Dell'Antonio:1991xt}. Then, since all physical configurations have at least one representative inside the Gribov region, the restriction is a consistent improvement of the Faddeev-Popov trick. The restriction of the path integral to the Gribov region implies on a dramatic modification of the gluon and ghost propagators. In one hand, the gluon propagator is suppressed at the infrared regime and acquires imaginary poles, on the other hand, the ghost propagator is enhanced by an infrared behaviour of the type $1/k^{4}$. These properties show, in an explicit way, that the elimination of the infinitesimal copies is of great importance for a consistent quantization, deeply modifying the theory and providing evidences of confinement.

The solution proposed by Gribov works nicely when the Faddeev-Popov operator is hermitian. The reason is that hermiticity ensures that the spectrum of the Faddeev-Popov operator is real and, therefore, it is possible to establish an order relation between the eigenvalues of such operator. Hence, it is possible to define a region where the Faddeev-Popov operator is positive-definite and the restriction of the path integral to this region ensures the absence of infinitesimal copies. However, if we desire to work with non-hermitean Faddeev-Popov operators, to perform such restriction is not a clear procedure because the order relation and, therefore, the definition of a region, do not make sense anymore. In this sense, hermiticity of the Faddeev-Popov operator plays a fundamental role for the elimination of copies \emph{\`a la} Gribov. 

After the work of Gribov, D.~Zwanziger implemented the restriction of the path integral to the Gribov region by the introduction of a nonlocal term to the original action \cite{Zwanziger:1988jt,Zwanziger:1989mf,Zwanziger:1992qr}. This term is commomnly known as \emph{horizon function}. The horizon function is an all order generalization of the Gribov approximated restriction term. This term can be localized by the introduction of a set of auxiliary fields and the final action is known as \emph{Gribov-Zwanziger action} \cite{Zwanziger:1988jt,Zwanziger:1989mf,Zwanziger:1992qr}, which is renormalizable at least to all orders in perturbation theory \cite{Zwanziger:1989mf,Maggiore:1993wq}. A proof of equivalence between Gribov (restriction through the ghost propagator) and Zwanziger (restriction through the horizon function) approaches, at the Landau gauge, can be found in \cite{Capri:2012wx}.

The Gribov-Zwanziger action, although free of infinitesimal copies, does not provide results for the gluon and ghost propagators fully compatible with the recent lattice results \cite{Cucchieri:2007md,Cucchieri:2009zt}. In order to improve these results, more non-perturbative effects, such as condensates, were considered. The result of the inclusion of such condensates to the Gribov-Zwanziger action is the well-known \emph{refined Gribov-Zwanziger action} (RGZ) which leads naturally to gluon and ghost propagators in remarkable harmony with the lattice results \cite{Dudal:2008sp,Dudal:2011gd}.

Gribov copies and the Gribov-Zwanziger framework were firstly studied at the Landau gauge. However, many results were also obtained at the maximal Abelian gauge (MAG), see for instance \cite{Capri:2006cz,Capri:2010an} and references therein. Essentially, these are the two covariant gauges where infinitesimal Gribov copies are, until now, reasonably understood. The MAG is known as a very important gauge for non-perturbative studies in the context of the dual superconductivity model for confinement \cite{'tHooft:1981ht}. Due to its decomposition in diagonal and off-diagonal components, the MAG is very appropriate to the study of the so-called Abelian dominance \cite{Ezawa:1982bf}.

Alternatively to the methods of Gribov and Zwanziger on the elimination of infinitesimal Gribov copies, the authors have developed a relatively simple new method to account for the Gribov ambiguities, at least at infinitesimal level and a restricted class of gauges, see \cite{Pereira:2013aza}. In this approach, the zero-modes equation for the Faddeev-Popov is purposely ruined in order to avoid Gribov copies. Then, the ruined equation is implemented at the gauge fixed action as an extra constraint. Therefore, all infinitesimal Gribov copies are eliminated at the classical level. In a certain sense, this elimination is direct and does not require the construction of a geometric region to restrict the path integral. The only requirement is to avoid all zero-modes, which characterize the infinitesimal copies. Since the identification of Gribov copies and the zero-modes of the Faddeev-Popov operator is independent of the hermiticity of this operator, this method should also be employable to treat gauges with non-hermitian Faddeev-Popov operators. Thus, this method brings a new perspective on the elimination of copies. It is important to recall that, the method developed in \cite{Pereira:2013aza} requires exclusively $A$-dependent gauge conditions, although any particular choice was imposed. Therefore, in principle, there is a large class of gauges for which the method would be applicable. In particular, still in \cite{Pereira:2013aza}, consistency tests were made by applying the method to the Landau and maximal Abelian gauges. It is worth mention that the method here described is not the only alternative to the Gribov and Zwanziger techniques. There are other alternative techniques to deal with the Gribov ambiguities, see for instance \cite{Serreau:2012cg,Serreau:2013ila}.

In the present work, we apply this new method \cite{Pereira:2013aza} to a gauge with non-hermitian Faddeev-Popov operator: the Landau - maximal Abelian interpolating gauge (LMAIG) \cite{Dudal:2005zr,Capri:2005zj,Capri:2006bj}. This gauge has, at least, three advantages to motivate the present investigation. The first one, as already mentioned, is that the traditional approaches are not able, in principle, to deal with this gauge or any other gauge with non-hermitian Faddeev-Popov operators. Second, it is a gauge that link the two gauges where the Gribov problem can be handled. Thus, it is possible to verify the consistency of the results by interpolating among both limits of the LMAIG. Third, this gauge can be defined through a minimizing function given by
\begin{equation}
\mathcal{F} = \frac{1}{2}\int d^{4}x\;(A^{a}_{\mu}A^{a}_{\mu} + \eta A^{i}_{\mu}A^{i}_{\mu})\;,
\label{mfunctional}
\end{equation}
where $\eta$ is the interpolating parameter and the indices refer to the non-Abelian and Abelian sectors of the $SU(N)$ group (see Sec.~\ref{lmagintgauge} for the conventions). The gauge conditions of the LMAIG can be obtained by the minimization of the operator \eqref{mfunctional} with respect to gauge transformations. This means that the LMAIG could be, in principle, implemented on the lattice. This is a very welcome feature because it can work as a test for the application of the method\footnote{It is a well established fact that lattice techniques are the most trustful non-perturbative method to study Yang-Mills theories.}.

This paper is organized as follows: in Sect.~\ref{method}, a brief review of the method developed in \cite{Pereira:2013aza} is given. In Sect.~\ref{lmagintgauge}, we provide a review of the decomposition into diagonal and off-diagonal components of algebra-valued quantities, present the maximal Abelian gauge and make the explicit decomposition of the Landau gauge. After this, we introduce the LMAIG and discuss its important features for the analysis of Gribov copies. Then, in Sect.~4, we apply the method to the LMAIG, and construct an action free of infinitesimal copies. This is done in Sect.~\ref{methodapp}. In Sect.~\ref{propagatorsgluon} we calculate the diagonal and off-diagonal gluon propagators and show how it is possible to deform it into Landau and maximal Abelian gauges propagators. In Sect.~\ref{GAP} we make some comments about the gap equation in this method. Finally, in Sect.~7, we provide our conclusions. Many algebraic details were left to appendices to avoid big interruptions along the text.

\section{A brief review of the method} \label{method}

The elimination method proposed in \cite{Pereira:2013aza} is based on the introduction of an extra constraint that ruins the Gribov copies equation. In this section, we provide a brief review of the method in order to apply it to the interpolating LMAIG \cite{Capri:2005zj,Capri:2006bj}. It is not our intent to be rigorous here. For any formal detail we refer to \cite{Pereira:2013aza}.

Let us consider Yang-Mills theory for a given semi-simple Lie group $G$. We choose a gauge condition $\Delta^A$ that depends exclusively on the gauge field, \textit{i.e.} $\Delta^A = \Delta^A(A)$, where the group indices vary as $A,B,C,\ldots\in\{1,2,\ldots,\dim G\}$. As pointed out by Gribov \cite{Gribov:1977wm}, the Faddeev-Popov gauge fixing procedure does not ruin completely the gauge symmetry. Thus, some redundant configurations, which are connected by gauge transformations, are still being considered in the path integral. The existence of these copies depends on the existence of solutions for the Gribov copies equation, and this is obtained by the requirement of gauge invariance of the gauge condition, \textit{i.e.}
\begin{equation}
\Delta(A) = 0 \;\;\;\; \Rightarrow \;\;\;\; \Delta\left(A'=\frac{1}{g}U^{-1}\mathrm{d}U + U^{-1}AU\right)=0\;,
\label{copieseq}
\end{equation}
where $U\in G$ and $g$ is the coupling parameter. Besides the fact that we do not have much knowledge about the elimination of Gribov copies generated by large gauge transformations, we reasonably understand how to handle those copies generated by infinitesimal transformations. For this reason, we restrict ourselves to this case. We must warn the reader that, this method does not provide a full solution to the Gribov problem, since copies generated by large gauge transformations are not taken into account. However, the elimination of infinitesimal copies already gives very important modifications on the theory and defines a renormalizable local action (at least for the known Landau and maximal Abelian gauges) which justifies its study. The infinitesimal gauge transformation is then given by
\begin{equation}
\delta_{g}A^{A}_{\mu} = D^{AB}_{\mu}\zeta^{B}\;,\label{gaugetrans}
\end{equation}
where $\zeta^{B}$ is the infinitesimal gauge parameter. Thus, the copies equation \eqref{copieseq} becomes
\begin{equation}
\Delta(A^{A}_{\mu} + D^{AB}_{\mu}\zeta^{b}) = 0\;,\label{infcopieseq}
\end{equation}
where $D^{AB}_{\mu} \equiv \delta^{AB}\partial_{\mu} - gf^{ABC}A^{C}_{\mu}$  is the covariant derivative and $f^{ABC}$ represents the structure constants. At first order in $\zeta$, Eq.~\eqref{infcopieseq} can be written as
\begin{equation}
\nabla^{AB}\zeta^{B}=0\;,\label{infcopieseqFP}
\end{equation}
where $\nabla^{AB}$ is the Faddeev-Popov operator,
\begin{equation}
\nabla^{AB}=\frac{\partial \Delta^{A}}{\partial A^{C}_{\mu}}D^{CB}_{\mu}\;.\label{infcopieseqFP2}
\end{equation}
Summarizing, Eq.~\eqref{infcopieseqFP} is obtained by requiring infinitesimal gauge invariance of $\Delta(A)$.

The BRST transformation defined through the nilpotent operator $s$ is given by
\begin{eqnarray}
sA^{A}_{\mu} &=& -D^{AB}_{\mu}c^{B}\;, \nonumber \\
sc^A &=& \frac{g}{2}f^{ABC}c^Bc^C\;, \nonumber \\
s\overline{c}^A&=&ib^A\;,\nonumber\\
sb^A&=&0\;,
\label{brsttranstot}
\end{eqnarray}
where $c^A$ is the Faddeev-Popov ghost field, $\overline{c}^{A}$ is the antighost field and $b^{A}$ is the auxiliary Nakanishi-Lautrup field. It is immediate to see that the first equation of \eqref{brsttranstot} has the same form of \eqref{gaugetrans}. Of course, we have to understand that these are different transformations: The BRST, in particular, transforms a field with vanishing ghost number into a composite field with ghost number +1, while the gauge transformation does not change the ghost number. Nevertheless, it was proved in \cite{Pereira:2013aza} that these transformations are homotopic. Thus, since they have the same formal structure, we can obtain the copies equation by requiring, not the gauge invariance of the gauge condition, but the \textit{BRST invariance}\footnote{This is a particular property of gauge conditions which depend exclusively on the gauge field $A$, because of the very same formal structure between the gauge and BRST transformations.} of the very same gauge condition. Hence, we can write the copies equation as
\begin{equation}
s\Delta^{A}(A) = 0 \;\;\;\; \Rightarrow \;\;\;\; \nabla^{AB}c^{B} = 0\;.
\label{copieseqbrs}
\end{equation}

The key point of the method resides at this stage: Since we want a theory free of copies, we have to ruin the copies equation. This might be seen as a new constraint for the theory. Thus, from Eq.~\eqref{copieseqbrs}, we can see that, to ruin this equation, we need to break the BRST invariance of the copies equation. In this sense, we want to write an equation such that
\begin{equation}
\nabla^{AB}c^{B} = \Omega^{A}\;,
\label{noncopieseq}
\end{equation} 
where the term $\Omega^{A}$ must prevent the theory to develop infinitesimal copies. Roughly speaking, this is the main idea behind the method. Now, in order to implement Eq.~\eqref{noncopieseq} in a gauge theory, we have to be careful to preserve all well established features of the perturbative regime. A very important requirement we have to make is that the BRST symmetry is restored at the perturbative regime. This means that the BRST breaking must be \textit{soft}. Another requirement is that, since we do not want to affect the ghost sector, which is of great importance for the perturbative sector, we must introduce a set of trivial auxiliary fields to mimic Eq.~\eqref{copieseqbrs}. Finally, to ruin the copies equation and impose a consistent equation compatible with \eqref{noncopieseq}, the introduction of a soft BRST breaking term must be performed. As argued in \cite{Pereira:2013aza}, these goals are achieved by the introduction of two extra terms to the gauge fixed action, namely $S_{triv}$ and $\Xi$. These terms are responsible to implement a new constraint to the theory, satisfying the requirements mentioned before and reproducing Eq.~\eqref{noncopieseq}. Thus, we impose the action
\begin{equation}
S = S_{\mathrm{YM}} + S_{gf}+  S_{triv} + \Xi\;,
\label{fullactiongribov}
\end{equation}
where
\begin{equation}
S_{\mathrm{YM}} = \frac{1}{4}\int d^{4}x\;F^{A}_{\mu \nu}F^{A}_{\mu \nu} \;\;\;\; \mathrm{and} \;\;\;\; S_{gf} = \int d^{4}x\; \left(ib^{A}\Delta_{A} - \overline{c}^{A}s\Delta_{A}\right)\;.
\label{ymegf}
\end{equation}
As stated before, the term $S_{triv}$ is introduced to mimic the copies equation. To do so, we introduce a BRST quartet
\begin{eqnarray}
s\overline{\omega}^{AB}_\mu&=&\overline{\varphi}^{AB}_\mu\;,\nonumber\\
s\overline{\varphi}^{AB}_\mu&=&0\;, \nonumber \\
s\varphi^{AB}_\mu&=&\omega^{AB}_\mu\;,\nonumber\\
s\omega^{AB}_\mu&=&0\;, 
\label{brs6}
\end{eqnarray}
in such a way that
\begin{eqnarray}
S_{triv}&=& s\int d^4x\;\overline{\omega}^{AC}_\mu\nabla^{AB} \varphi^{BC}_\mu\; \nonumber\\
&=&  \int d^4x\left[\overline{\varphi}^{AC}_\mu\nabla^{AB} \varphi^{BC}_\mu-\overline{\omega}^{AC}_\mu\nabla^{AB} \omega^{BC}_\mu+\overline{\omega}^{AC}_\mu \left(D^{DE}_{\nu}c^E\right)\frac{\delta\nabla^{AB}}{\delta A_\nu^D}\varphi^{BC}_\mu\right]\;.\label{gz0}
\end{eqnarray}
It is easy to see that the equation of motion for $\overline{\varphi}$ produced by $S_{triv}$ is exactly the copies equation. Moreover, the indices of the auxiliary fields are not arbitrary and describe the degeneracy of the copies equation. Since our point is precisely to ruin this equation, the term $\Xi$ has the following general form
\begin{eqnarray}
\Xi &=& \int d^4x\; \gamma^2D_\mu^{AB}(\varphi+\overline{\varphi})^{AB}_\mu+ \int d^4x\;\gamma^2\zeta_1\left(\overline{\varphi}^{AB}_\mu\varphi^{AB}_\mu-\overline{\omega}^{AB}_\mu\omega^{AB}_\mu\right) \nonumber \\
&+& \int d^4x\; \gamma^2\left(\zeta_2A_\mu^AA_\mu^A+\zeta_3\overline{c}^Ac^A\right) + \int d^4x\; \varepsilon\gamma^z \;,
\label{generalbreakingterm}
\end{eqnarray}
where $\gamma$ is a mass parameter introduced to fullfill the \textit{soft} breaking requirement. With $\Xi$, we see that the equation of motion for $\overline{\varphi}$ is modified and represents a ``ruined" copies equation
\begin{equation}
\nabla^{AB}\varphi^{BC}_{\mu} + \frac{\delta \Xi}{\delta \overline{\varphi}^{AC}_{\mu}} = 0\;,\label{ruinedcopieseq}
\end{equation}
which is the extra constraint that ensures the absence of infinitesimal copies.

The action given by Eq.~\eqref{fullactiongribov} is then an action which satisfies the constraint given by Eq.~(\ref{ruinedcopieseq}). With this we eliminate all infinitesimal copies at the classical level. This result qualitatively coincides with the well establish refined Gribov-Zwanziger action, see \cite{Dudal:2008sp}. It is worth mention that the form of the breaking term defined by eq.~(\ref{generalbreakingterm}) has a kind of ``freedom''. To ruin the copies equation we must add a term which will be responsible for the breaking of BRST invariance of the gauge condition. This term must depend on $\overline{\varphi}$ for the obvious reason that, if it does not, the variation of the action \eqref{fullactiongribov} with respect to $\overline{\varphi}$ would not produce a ``ruined" copies equation, as required. Moreover, the derivative of this term with respect to $\overline{\varphi}$ must depend exclusively on the gauge field $A$. The reason is that, if it depends on other fields, this term vanishes at their trivial vacua. Requiring the exclusive dependence on $A$, we ensure that the only copies that can be generated are related to $A=0$. However, if they exist, they are necessarily different from zero and, therefore, the constraint will eliminate the copy $A=0$ for the appropriate $A$. The conclusion is that the first term of Eq.(\ref{generalbreakingterm}),
\begin{equation}
\tilde{\Xi} = \int d^4x\;\gamma^2D_\mu^{AB}(\varphi+\overline{\varphi})^{AB}_\mu\;,
\label{xitilde}
\end{equation}
is sufficient for our requirements. In this sense, to ruin the copies equation in a minimal way, we could add only \eqref{xitilde} to the original action and it will generate a theory free of infinitesimal copies. It this case, the extra terms can be included by the LCO technique, in the usual way \cite{Dudal:2003pe,Verschelde:2001ia,Dudal:2002pq,Knecht:2001cc,Dudal:2003vv}. On the other hand, once $\gamma$ is at our disposal, the extra soft terms in \eqref{generalbreakingterm} are allowed by power counting. What would decide if they are present or not are the Ward identities of the particular chosen gauge. In both cases, the effect is the obtention of the refined Gribov-Zwanziger action \cite{Dudal:2008sp,Dudal:2011gd}. Furthermore, there is another possible freedom, for each term proportional to the mass parameter $\gamma$, we could replace it by different mass parameters. Essentially, this can also be obtained by the redefinition $m_i=\zeta_i\gamma$, which means that the independent character of these coefficients are accounted by the parameters $\zeta_i$. Let us also comment on the term proportional to $\zeta_1$. As it is possible to see from eq.(\ref{generalbreakingterm}), a larger field combination is associated with the parameter $\zeta_{1}$. The reason is that we can introduce such combination as a BRST exact form, $\gamma^{2}s\int d^{4}x\;\overline{\omega}^{AB}_{\mu}\varphi^{AB}_{\mu}$, for instance. We could introduce this mass terms in a independent way, but following the idea of a minimal breaking of the BRST symmetry, a BRST exact term fits better for our plans.

Finally, it is important to understand that the inclusion of all extra terms proportional to $\gamma$ implies on a deep modification of the so-called gap equation \cite{Pereira:2013aza}. Until now, no results are known for this generalized gap equation and we are not able to decide if it is a better choice to follow it or not. However, we will keep these terms as in \eqref{generalbreakingterm} because they are important to reproduce the refined Gribov-Zwanziger features and also because we will not deal with the gap equation in this work\footnote{The study of the alternative gap equation is left for future investigation. Probably starting with the Landau gauge case.}. In fact, if the extra terms are not allowed for any reason, all we have to do is to set the corresponding $\zeta_i$ to zero.

\section{The Landau and maximal Abelian gauges and their interpolation} \label{lmagintgauge}

From now on, we restrict ourselves to the $SU(N)$ gauge group. In \cite{Capri:2005zj,Capri:2006bj}, a gauge fixing which interpolates among Landau, Coulomb and maximal Abelian gauges was studied. In the present work, we will analyze the Gribov problem in this gauge. However, we will consider only the interpolation between Landau and the maximal Abelian gauges and avoid the Coulomb sector of the gauge. Since the MAG is characterized by imposing different gauge conditions to the diagonal and off-diagonal components of the Lie algebra-valued fields, we will decompose the Landau gauge in order to provide an explicit comparison with the reduction of the interpolating gauge to the Landau case\footnote{This step makes easier the comparison between the usual Landau gauge and the Landau limit of the interpolating gauge.}. To fix notation and conventions, we will briefly review this kind of decomposition, called Abelian decomposition \cite{'tHooft:1981ht}. Essentially, the $SU(N)$ group is dismembered into its Abelian and non-Abelian sectors where the Abelian sector is recognized as the Cartan subgroup. The gauge field decomposition is taken as
\begin{equation}
A_{\mu} = A_{\mu}^{A}G^{A} = A_{\mu}^{a}G^{a} + A_{\mu}^{i}G^{i}\;,
\label{adecomp}
\end{equation}
where $G^A$ correspond to the $(N^{2} - 1)$ generators of the $SU(N)$ group; $G^{a}$ are the $N(N-1)$ off-diagonal generators of the gauge group; and $G^i$ represent the $(N-1)$ Cartan subgroup generators. The indices $\left\{a,b,c,\ldots,h\right\}$ run from $1\; \mathrm{to}\; N(N-1)$ and the indices $\left\{i,j,k,\ldots\right\}$ run from $1\; \mathrm{to}\; (N-1)$. As a consequence of this decomposition, we can write the decomposed BRST transformations \eqref{brsttranstot} as
\begin{eqnarray}
sA_{\mu}^{a} &=& -(D_{\mu}^{ab}c^{b} + gf^{abc}A_{\mu}^{b}c^{c} + gf^{abi}A_{\mu}^{b}c^{i})\;, \nonumber \\
sc^{a} &=& gf^{abi}c^{b}c^{i} + \frac{g}{2}f^{abc}c^{b}c^{c}\;, \nonumber \\
s\overline{c}^{a} &=& ib^{a}\;, \nonumber \\
sb^{a} &=& 0\;, \nonumber \\
sA_{\mu}^{i} &=& -(\partial_{\mu}c^{i} + gf^{abi}A_{\mu}^{a}c^{b})\;, \nonumber \\
sc^{i} &=& \frac{g}{2}f^{abi}c^{a}c^{b}\;, \nonumber \\
s\overline{c}^{i} &=& ib^{i}\;, \nonumber \\
sb^{i} &=& 0\;,
\label{brstabelian}
\end{eqnarray}
where the covariant derivative is defined with respect to the Abelian sector and acts on non-Abelian quantities,
\begin{equation}
D^{ab}_{\mu} = \delta^{ab}\partial_{\mu} - gf^{abi}A^{i}_{\mu}\;.
\label{cderivative}
\end{equation}
We can now write the gauge fixed Yang-Mills action \eqref{ymegf} as
\begin{eqnarray}
S_{0} &=& S_{\mathrm{YM}} + S_{gf} + S_{\mathrm{ext}} \nonumber \\
&=& \int d^{4}x\; (F^{a}_{\mu \nu}F^{a}_{\mu \nu} + F^{i}_{\mu \nu}F^{i}_{\mu \nu}) + s\int d^{4}x\; (\overline{c}^{a}\Delta^{a} + \overline{c}^{i}\Delta^{i}) + S_{\mathrm{ext}},
\label{action1}
\end{eqnarray}
where $F^{a}_{\mu \nu}$ and $F^{i}_{\mu \nu}$ are the components of the field strength, which are, explicitly,

\begin{eqnarray}
F_{\mu \nu}^{a} &=& D_{\mu}^{ab}A_{\nu}^{b} - D_{\nu}^{ab}A_{\mu}^{b} + gf^{abc}A_{\mu}^{b}A_{\nu}^{c}, \nonumber \\
F_{\mu \nu}^{i} &=& \partial_{\mu}A_{\nu}^{i} - \partial_{\nu}A_{\mu}^{i} + gf^{abi}A_{\mu}^{a}A_{\nu}^{b}.
\label{Fcomponents}
\end{eqnarray}
and $\Delta^{a}(A)$ and $\Delta^{i}(A)$ are related to the gauge condition components. To complete the Abelian decomposition we can write the Jacobi identities as
\begin{eqnarray}
f^{abi}f^{bcj} + f^{abj}f^{bic} &=& 0, \nonumber \\
f^{abc}f^{cdi} + f^{adc}f^{cib} + f^{aic}f^{cbd} &=& 0, \nonumber \\
f^{abc}f^{cde} + f^{abi}f^{ide} + f^{adc}f^{ceb} + f^{adi}f^{ieb} + f^{aec}f^{cbd} + f^{aei}f^{ibd} &=& 0\;.
\label{jacobimag}
\end{eqnarray}

\subsection{The maximal Abelian gauge} \label{MAG}

The maximal Abelian gauge imposes different gauge conditions to the diagonal and off-diagonal sectors of the gauge fields, namely
\begin{eqnarray}
D_{\mu}^{ab}A_{\mu}^{b}&=&0\;, \nonumber \\ 
\partial_{\mu}A_{\mu}^i&=&0\;,
\label{magcond}
\end{eqnarray}
and the corresponding gauge fixing action is given by
\begin{eqnarray}
S_{\mathrm{MAG}} &=& \int d^4x \left[ib^{a}D^{ab}_{\mu}A^{b}_{\mu} + \overline{c}^{a}\nabla^{ab}c^{b} - gf^{abc}(D^{ad}_{\mu}A^{d}_{\mu})\overline{c}^{b}c^{c} 
- gf^{abi}(D^{ac}_{\mu}A^{c}_{\mu})\overline{c}^{b}c^{i} \right.\nonumber \\ 
&+& \left.ib^{i}\partial_{\mu}A^{i}_{\mu} + \overline{c}^{i}\partial_{\mu}(\partial_{\mu}c^{i} + gf^{abi}A^{a}_{\mu}c^{b})\right]\;,
\label{maggf}
\end{eqnarray}
where the operator $\nabla^{ab}$ is the Faddeev-Popov operator,
\begin{equation}
\nabla^{ab} = -D_{\mu}^{ac}D_{\mu}^{cb} - gf^{acd}A_{\mu}^{c}D_{\mu}^{db} - g^{2}f^{aci}f^{bdi}A_{\mu}^{c}A_{\mu}^{d}\;.
\label{fpmag}
\end{equation}
The gauge conditions \eqref{magcond} can be obtained from the equations of $b^a$ and $b^i$. If we think of the conditions for the existence of Gribov copies, we can derive the copies equation requiring the gauge/BRST invariance of the gauge condition \cite{Pereira:2013aza}. Hence, since we have two different gauge conditions in MAG, it is natural to expect two copies equations. In fact, if we calculate directly the copies equation from the gauge conditions \eqref{magcond}, we obtain the following equations
\begin{eqnarray}
\nabla^{ab}\zeta^{b} &=& 0\;, \nonumber \\
\partial_{\mu}(\partial_{\mu}\zeta^{i} + gf^{abi}A^{a}_{\mu}\zeta^{b}) &=& 0\;,
\label{copieseqnmag}
\end{eqnarray}
where $\zeta^{a}$ and $\zeta^{i}$ are the off-diagonal and diagonal components of the infinitesimal gauge parameter, respectively. From \eqref{copieseqnmag} we see that the first equation just involves the off-diagonal component of the gauge parameter while the second involves both. Simple manipulations of the second equation provide
\begin{equation}
\zeta^{i} = \frac{-gf^{abi}\partial_{\mu}(A^{a}_{\mu}\zeta^{b})}{\partial^{2}}\;,
\label{redundant}
\end{equation}
which shows that, once one has solved the first equation of \eqref{copieseqnmag}, the second does not contribute with any extra information. This redundancy is the reason why only the first equation of \eqref{copieseqnmag} is considered as the copies equation for the MAG. A final comment is that the Faddeev-Popov operator is hermitian in this case, see \cite{Capri:2006cz,Capri:2010an} and references therein for more details.

\subsection{The decomposed Landau gauge} \label{dlandau}

The Landau gauge condition
\begin{equation}
\Delta^{A} = \partial_{\mu}A^{A}_{\mu}=0\;,
\label{landaucondition}
\end{equation}
does not distinct the diagonal and off-diagonal sectors of the gauge connection. However, since we will work with decomposed fields, we also write the Landau gauge fixing relevant expressions in the Abelian decomposition. The result is the decomposed Landau gauge fixing action, given by
\begin{eqnarray}
S_{\mathrm{L}} &=& s \int d^4x (\overline{c}^{a}\partial_{\mu}A_{\mu}^{a} + \overline{c}^{i}\partial_{\mu}A_{\mu}^{i}) \nonumber \\
&=& \int \left[ib^{a}\partial_{\mu}A_{\mu}^{a} + \overline{c}^{a}\partial_{\mu}(D^{ab}_{\mu}c^{b} + gf^{abc}A^{b}_{\mu}c^{c} + gf^{abi}A_{\mu}^{b}c^{i}) + ib^{i}\partial_{\mu}A_{\mu}^{i} \right.\nonumber\\ 
&+& \left.\overline{c}^{i}\partial_{\mu}(\partial_{\mu}c^{i} + gf^{abi}A_{\mu}^{a}c^{b})\right]\;.
\label{gflandau}
\end{eqnarray}
It is immediate to obtain the Faddeev-Popov operator from \eqref{gflandau}. In components, it is given by
\begin{eqnarray}
\nabla^{ab} &=& - \partial_{\mu}D_{\mu}^{ab} + gf^{abc}A^{c}_{\mu}\partial_{\mu}\;, \nonumber \\
\nabla^{ai} &=& - gf^{abi}A_{\mu}^{b}\partial_{\mu}\;, \nonumber \\
\nabla^{ia} &=& gf^{abi}A^{b}_{\mu}\partial_{\mu}\;, \nonumber \\
\nabla^{ij} &=& - \delta^{ij}\partial^{2}\;. 
\label{FPlandau}
\end{eqnarray}
Unlike the case of MAG, all components above in \eqref{FPlandau} contribute to the copies equations. If we again consider the diagonal and off-diagonal components of the infinitesimal gauge parameter, $\zeta^{i}$ and $\zeta^{a}$, respectively, we can write the following copies equations
\begin{eqnarray}
\nabla^{ab}\zeta^{b} + \nabla^{ai}\zeta^{i} &=& 0\;, \nonumber \\
\nabla^{ia}\zeta^{a} + \nabla^{ij}\zeta^{j} &=& 0\;.\label{gcx}
\end{eqnarray}
In this case, both equations encompass all components of the infinitesimal gauge parameter. Hence, we cannot put away any of them and all components of the Faddeev-Popov operator are essential to the analysis. 

Another important remark is that the full Faddeev-Popov operator for the Landau gauge is also hermitian, see for instance \cite{Gribov:1977wm,Sobreiro:2005ec}. If we adopt a matrix viewpoint, an hermitian operator is such that the elements of its diagonal are hermitian operators and all elements above the diagonal are hermitian conjugate of the elements below it. If we analyze the mixed components of \eqref{FPlandau} we can see that $(\nabla^{ai})^{T \ast} = \nabla^{ia}$.

\subsection{Interpolating gauge}

In order to provide a gauge fixing which interpolates among Landau and maximal Abelian gauges \cite{Capri:2005zj}, we introduce a real interpolating parameter $\eta$ and write the following gauge conditions
\begin{eqnarray}
D^{ab}_{\mu}A^{b}_{\mu} + \eta f^{abi}A^{i}_{\mu}A^{b}_{\mu} &=& 0\;, \nonumber \\
\partial_{\mu}A^{i}_{\mu} &=& 0\;.
\label{intgcond}
\end{eqnarray}
Thus, it is clear that the gauge condition for the diagonal component of the gauge field is identical for the Landau and maximal Abelian gauges cases. Moreover, for the first equation of \eqref{intgcond}, the case $\eta = 1$ gives the Landau gauge condition while for $\eta = 0$, the MAG condition is achieved. Consequently, we can write the gauge fixing term as
\begin{eqnarray}
S_{\mathrm{LM}} &=& s\int d^{4}x \left(\overline{c}^{a}D^{ab}_{\mu}A^{b}_{\mu} + \eta g\overline{c}^{a}f^{abi}A^{i}_{\mu}A^{b}_{\mu} + \overline{c}^{i}\partial_{\mu}A^{i}_{\mu}\right) \nonumber \\
&=& \int d^{4}x \left(b^{a}D^{ab}_{\mu}A^{b}_{\mu} + \overline{c}^{a}D^{ab}_{\mu}D^{bc}_{\mu}c^{c} + g\overline{c}^{a}f^{abi}(D^{bc}_{\mu}A^{c}_{\mu})c^{i} + g\overline{c}^{a}D^{ab}_{\mu}(f^{bcd}A^{c}_{\mu}c^{d}) \right.\nonumber \\
&-& \left.g^{2}f^{abi}f^{cdi}\overline{c}^{a}c^{d}A^{b}_{\mu}A^{c}_{\mu} + b^{i}\partial_{\mu}A^{i}_{\mu} + \overline{c}^{i}\partial_{\mu}(\partial_{\mu}c^{i} + gf^{iab}A^{a}_{\mu}c^{b}) + \eta gf^{abi}A^{a}_{\mu}(\partial_{\mu}c^{i})\overline{c}^{b}  \right.\nonumber \\
&+& \left.\eta g^{2}f^{abi}f^{cdi}\overline{c}^{a}c^{d}A^{b}_{\mu}A^{c}_{\mu}- \eta gf^{abi}A^{i}_{\mu}A^{a}_{\mu}(b^{b} - gf^{bcj}\overline{c}^{c}c^{j}) + \eta gf^{abi}A^{i}_{\mu}(D^{ac}_{\mu}c^{c})\overline{c}^{b}  \right.\nonumber \\
&+& \left.\eta g^{2}f^{abi}f^{acd}A^{i}_{\mu}A^{c}_{\mu}c^{d}\overline{c}^{b}\right)\;.
\label{gfix}
\end{eqnarray}
It is a simple exercise to verify that for $\eta = 0$, Eq.~\eqref{gfix} reduces to Eq.~\eqref{maggf}, and for $\eta = 1$, it reduces to Eq.~\eqref{gflandau}.

\section{Eliminating the Gribov copies} \label{methodapp}

\subsection{The Faddeev-Popov operator and Gribov ambiguities} \label{fpoperatorinter}

The gauge conditions presented in the last Section provide a way to interpolate among the Landau and maximal Abelian gauges. To analyze the Gribov problem in this gauge, it is fundamental to study the Faddeev-Popov operator in order to establish the copies equation and its main properties. The way we do this is completely analogous to the MAG (as showed in Sect.~\ref{MAG}). Thus, by requiring gauge/BRST invariance of \eqref{intgcond}, the following operators are obtained
\begin{eqnarray}
\nabla^{ab} &=& -D^{ac}_{\mu}D^{cb}_{\mu} - gf^{acd}A^{c}_{\mu}(D^{db}_{\mu} + g\eta f^{dbi}A^{i}_{\mu}) + \eta gf^{cai}A^{i}_{\mu}D^{cb}_{\mu} + (1-\eta)g^{2}f^{adi}f^{cbi}A^{d}_{\mu}A^{c}_{\mu}\;, \nonumber \\
\nabla^{ai} &=& - \eta gf^{abi}A^{b}_{\mu}\partial_{\mu}\;, \nonumber \\
\nabla^{ia} &=& gf^{abi}(\partial_{\mu}A^{b}_{\mu} + A^{b}_{\mu}\partial_{\mu})\;, \nonumber \\
\nabla^{ij} &=& - \delta^{ij} \partial^{2}\;.
\label{fpinterpolante}
\end{eqnarray}
These operators act on a gauge parameter pair $(\zeta^a,\zeta^i)$, exactly as in \eqref{gcx}. First of all, as a consistency check, we must verify if, for the suitable choices of the parameter $\eta$, this operator reduces to the previous operators for Landau and maximal Abelian gauges. Hence, starting with $\eta = 0$, the first equation of \eqref{fpinterpolante} reduces immediately to \eqref{fpmag}, the second turns out to be the null operator and the last remain unaffected. Actually, the two last equations simply define the redundant condition \eqref{redundant} for the MAG. We conclude then that the choice $\eta = 0$ returns the Faddeev-Popov operator of the MAG, as expected. On the other hand, if we choose $\eta = 1$, we can see that, after some simple manipulations, the first equation of \eqref{fpinterpolante} reduces to $\nabla^{ab} = -\partial_{\mu}D^{ab}_{\mu} - gf^{acb}A^{c}_{\mu}\partial_{\mu}$, which is exactly the purely off-diagonal components of the Faddeev-Popov operator for the Landau gauge. The second equation of (\ref{fpinterpolante}) reduces to the second equation of (\ref{FPlandau}). The third equation of (\ref{fpinterpolante}) does not involve the interpolating parameter $\eta$, but once we choose $\eta = 1$, the first gauge condition of (\ref{intgcond}) turns out to be $\partial_{\mu}A^{a}_{\mu}=0$, which means that we can substitute this in the third equation of (\ref{fpinterpolante}) and obtain the same result for the Landau gauge (\ref{FPlandau}). Finally, the fourth equation is unchanged. Summarizing, these are the components of the Faddeev-Popov operator for the Landau gauge. This concludes our consistency checks for now. 

Let us make a quick remark about the Faddeev-Popov operator. It is widely known that, in standard techniques employed to deal with the Gribov problem \cite{Gribov:1977wm,Sobreiro:2005ec}, the hermiticity of the Faddeev-Popov operator is essential. Since we can associate Gribov copies with zero-modes of the Faddeev-Popov operator, the knowledge about its spectrum is very welcome. A hermitian operator has only real eigenvalues, allowing to establish an order relation between them. In the Landau gauge, for instance, is through this analysis that it is possible to construct a region where the Faddeev-Popov operator is positive-definite. For this reason, it is possible to eliminate all infinitesimal Gribov copies from the path integral by the restriction of the integration to this domain. This technique makes the analysis of the Gribov problem a geometrical problem. On the other hand, this method cannot be employed to non-hermitian Faddeev-Popov operators. It is then not clear how to generate a region that will restrict the path integral. Nevertheless, the method developed in \cite{Pereira:2013aza} and briefly reviewed in Sect.~\ref{method} does not require the definition of a region to perform the functional integration. Thus, we can apply it for gauges with non-hermitian Faddeev-Popov operators. In fact, as discussed in \cite{Pereira:2013aza}, in the case of hermitian Faddeev-Popov operators, the new method is equivalent to restrict the path integral to a region defined by the zero-modes of the corresponding FaddeevPopov operator.

Getting back to the LMAIG Faddeev-Popov operator \eqref{fpinterpolante}, its first decomposed operator can be rewritten as
\begin{equation}
\nabla^{ab} = \tilde{\nabla}^{ab} - g^{2}\eta f^{acd}f^{dbi}A^{c}_{\mu}A^{i}_{\mu} - \eta gf^{aci}A^{i}_{\mu}D^{cb}_{\mu} - \eta g^{2}f^{adi}f^{cbi}A^{d}_{\mu}A^{c}_{\mu}\;,
\label{nablianaint}
\end{equation}
where 
\begin{equation}
\tilde{\nabla}^{ab} = -D^{ac}_{\mu}D^{cb}_{\mu} - gf^{acd}A^{c}_{\mu}D^{db}_{\mu} - g^{2}f^{adi}f^{bci}A^{d}_{\mu}A^{c}_{\mu}\;,
\label{nablatilde}
\end{equation}
is the Faddeev-Popov operator for the MAG. As mentioned in Sect.~\ref{MAG}, the operator defined by Eq.~\eqref{nablatilde} is hermitian. It is possible to show that the operator $\nabla^{ab} - \tilde{\nabla}^{ab}$ is not hermitian, the details of the proof can be found at Ap.~\ref{herm}. The purely diagonal component of the Faddeev-Popov operator is trivially hermitian. Now, only the mixed components are left and we will follow an analogous idea presented in Sect.~\ref{dlandau}. The whole idea is based on the fact that we can write the Faddeev-Popov operator in a matrix form and it has two blocks formed by the purely off-diagonal and diagonal components. Terms outside these blocks are the mixed ones and to analyze their hermiticity we must take their hermitian conjugate. Thus, transposing and taking the complex conjugate of the matrix
\begin{equation}
(\nabla^{ai})^{\dagger} = - \eta gf^{iba}A^{b}_{\mu}\partial_{\mu} - \eta gf^{iba}\partial_{\mu}A^{b}_{\mu}= \eta gf^{abi}(A^{b}_{\mu}\partial_{\mu} + \partial_{\mu}A^{b}_{\mu})\;.
\label{nhermitean}
\end{equation}
it is clear that $(\nabla^{ai})^{\dagger} \ne \nabla^{ia}$. In fact, since one involves the parameter $\eta$ and the other does not, it is impossible to establish a hermiticity relation between these components. Thus, pictorially, we have that the full Faddeev-Popov operator,
\[ \nabla = \left( \begin{array}{ccc}
\nabla^{ab} & -g\eta f^{abi}A^{b}_{\mu}\partial_{\mu} \\
   \\
gf^{abi}(\partial_{\mu}A^{b}_{\mu} + A^{b}_{\mu}\partial_{\mu}) & \nabla^{ij} 
\end{array} \right)\;,\]
is not hermitian. Moreover, unlikely the MAG, it is not possible to eliminate some components of this operator to analyze the Gribov copies. See \eqref{gcx}. This, and the fact that the LMAIG is an exclusively $A$-dependent gauge, are the leads allowing the direct elimination of the zero-modes within the method developed in \cite{Pereira:2013aza}. 

\subsection{Trivial set of auxiliary fields}

According to the method, it is possible to eliminate Gribov copies directly, by imposing a new constraint to the theory. This constraint is, essentially, the requirement that the copies equation is not obeyed. This is done by the introduction of a set of auxiliary fields, forming a BRST quartet, through a trivial term and a soft BRST breaking term. From now on we will deal exclusively with the interpolating gauge, so when we refer to the Faddeev-Popov operator, we are talking about the operator defined by Eq.~\eqref{fpinterpolante}.

The trivial term is given by\footnote{Here, we introduced a global minus sign in Eq.~\eqref{triv} because of our definition of the Faddeev-Popov operator in Eq.~\eqref{fpinterpolante}.} 
\begin{eqnarray}
S_{triv} &=& - s\int d^{4}x~ \overline{\omega}^{AC}_{\mu}\nabla^{AB}\varphi^{BC}_{\mu} \nonumber \\
&=& - \int d^{4}x \left[\overline{\varphi}^{AC}_{\mu}\nabla^{AB}\varphi^{BC}_{\mu} - \overline{\omega}^{AC}_{\mu}\nabla^{AB}\varphi^{BC}_{\mu} - \overline{\omega}^{AC}_{\mu}(sA^{D}_{\nu})\frac{\delta \nabla^{AB}}{\delta A^{D}_{\nu}}\varphi^{BC}_{\mu}\right]\;,
\label{triv}
\end{eqnarray}
where capital latin indices refer to the complete Lie algebra. The full decomposition of \eqref{triv} in off-diagonal and diagonal components results in
\begin{eqnarray}
S_{triv} &=& \int d^{4}x \left[\overline{\varphi}^{ac}_{\mu}\nabla^{ab}\varphi^{bc}_{\mu} + \overline{\varphi}^{ac}_{\mu}\nabla^{ai}\varphi^{ic}_{\mu} + \overline{\varphi}^{aj}_{\mu}\nabla^{ab}\varphi^{bj}_{\mu} + \overline{\varphi}^{aj}_{\mu}\nabla^{ai}\varphi^{ij}_{\mu} + \overline{\varphi}^{ic}_{\mu}\nabla^{ib}\varphi^{bc}_{\mu} + \overline{\varphi}^{ic}_{\mu}\nabla^{ij}\varphi^{jc}_{\mu} \right.\nonumber \\ 
&+& \left.\overline{\varphi}^{ij}_{\mu}\nabla^{ib}\varphi^{bj}_{\mu} 
+ \overline{\varphi}^{ij}_{\mu}\nabla^{ik}\varphi^{kj}_{\mu} - \overline{\omega}^{ac}_{\mu}\nabla^{ab}\omega^{bc}_{\mu} - \overline{\omega}^{ac}_{\mu}\nabla^{ai}\omega^{ic}_{\mu} - \overline{\omega}^{aj}_{\mu}\nabla^{ab}\omega^{bj}_{\mu} - \overline{\omega}^{aj}_{\mu}\nabla^{ai}\omega^{ij}_{\mu} \right.\nonumber \\ 
&-& \left.\overline{\omega}^{ic}_{\mu}\nabla^{ib}\omega^{bc}_{\mu} 
-\overline{\omega}^{ic}_{\mu}\nabla^{ij}\omega^{jc}_{\mu} 
-\overline{\omega}^{ij}_{\mu}\nabla^{ib}\omega^{bj}_{\mu} - \overline{\omega}^{ij}_{\mu}\nabla^{ik}\omega^{kj}_{\mu} - \overline{\omega}_{\mu}^{ac}(sA^{d}_{\nu})\frac{\delta\nabla^{ab}}{\delta A^{d}_{\nu}}\varphi^{bc}_{\mu} \right.\nonumber \\ 
&-& \left.\overline{\omega}_{\mu}^{ac}(sA^{i}_{\nu})\frac{\delta\nabla^{ab}}{\delta A^{i}_{\nu}}\varphi^{bc}_{\mu} 
-\overline{\omega}_{\mu}^{ac}(sA^{d}_{\nu})\frac{\delta\nabla^{ai}}{\delta A^{d}_{\nu}}\varphi^{ic}_{\mu} 
- \overline{\omega}_{\mu}^{aj}(sA^{d}_{\nu})\frac{\delta\nabla^{ab}}{\delta A^{d}_{\nu}}\varphi^{bj}_{\mu} - \overline{\omega}_{\mu}^{aj}(sA^{i}_{\nu})\frac{\delta\nabla^{ab}}{\delta A^{i}_{\nu}}\varphi^{bj}_{\mu} \right.\nonumber \\ 
&-& \left.\overline{\omega}_{\mu}^{aj}(sA^{d}_{\nu})\frac{\delta\nabla^{ai}}{\delta A^{d}_{\nu}}\varphi^{ij}_{\mu}  
-\overline{\omega}_{\mu}^{ic}(sA^{d}_{\nu})\frac{\delta\nabla^{ib}}{\delta A^{d}_{\nu}}\varphi^{bc}_{\mu} 
- \overline{\omega}_{\mu}^{ij}(sA^{d}_{\nu})\frac{\delta\nabla^{ib}}{\delta A^{d}_{\nu}}\varphi^{bj}_{\mu} \right]\;, 
\label{trividec}
\end{eqnarray}
where terms involving the functional derivative with respect to $A$ of $\nabla^{ij}$ and terms involving the functional derivative with respect to $A^{j}$ of $\nabla^{ia}$ and $\nabla^{ai}$ are not present because they vanish. The explicit form of (\ref{trividec}) is
\begin{eqnarray}
S_{triv} &=& \int d^{4}x \left\{\overline{\varphi}^{ac}_{\mu}\nabla^{ab}\varphi^{bc}_{\mu} + \overline{\varphi}^{ac}_{\mu}\nabla^{ai}\varphi^{ic}_{\mu} + \overline{\varphi}^{aj}_{\mu}\nabla^{ab}\varphi^{bj}_{\mu} + \overline{\varphi}^{aj}_{\mu}\nabla^{ai}\varphi^{ij}_{\mu} + \overline{\varphi}^{ic}_{\mu}\nabla^{ib}\varphi^{bc}_{\mu} + \overline{\varphi}^{ic}_{\mu}\nabla^{ij}\varphi^{jc}_{\mu} \right.\nonumber \\ 
&+& \left.\overline{\varphi}^{ij}_{\mu}\nabla^{ib}\varphi^{bj}_{\mu} + \overline{\varphi}^{ij}_{\mu}\nabla^{ik}\varphi^{kj}_{\mu} - \overline{\omega}^{ac}_{\mu}\nabla^{ab}\omega^{bc}_{\mu} - \overline{\omega}^{ac}_{\mu}\nabla^{ai}\omega^{ic}_{\mu} - \overline{\omega}^{aj}_{\mu}\nabla^{ab}\omega^{bj}_{\mu} - \overline{\omega}^{aj}_{\mu}\nabla^{ai}\omega^{ij}_{\mu} \right.\nonumber \\ 
&-& \left.\overline{\omega}^{ic}_{\mu}\nabla^{ib}\omega^{bc}_{\mu} - \overline{\omega}^{ic}_{\mu}\nabla^{ij}\omega^{jc}_{\mu} 
- \left[gf^{bdi}(\partial_{\nu}\overline{\omega}^{ic}_{\mu})\varphi^{bc}_{\mu} + gf^{bdi}(\partial_{\nu}\overline{\omega}^{ij}_{\mu})\varphi^{bj}_{\mu} + \eta gf^{adi}\overline{\omega}^{ac}_{\mu}\partial_{\nu}\phi^{ic}_{\mu} \right.\right.\nonumber \\ 
&+& \left.\left.\eta gf^{adi}\overline{\omega}^{aj}_{\mu}\partial_{\nu}\varphi^{ij}_{\mu} - gf^{adb}\overline{\omega}^{ac}_{\mu}\partial_{\nu}\phi^{bc}_{\mu} 
+g^{2}(1-\eta)f^{ade}f^{ebi}\overline{\omega}^{ac}_{\mu}A^{i}_{\nu}\varphi^{bc}_{\mu} + g^{2}(1-\eta)(f^{adi}f^{ebi} \right.\right.\nonumber \\
&+& \left.\left.f^{aei}f^{dbi})\overline{\omega}^{ac}_{\mu}A^{e}_{\nu}\phi^{bc}_{\mu} - gf^{adb}\overline{\omega}^{aj}_{\mu}\partial_{\nu}\varphi^{bj}_{\mu} 
+ g^{2}(1-\eta)f^{ade}f^{ebi}\overline{\omega}^{aj}_{\mu}A^{i}_{\nu}\varphi^{bj}_{\mu} + g^{2}(1-\eta)(f^{adi}f^{ebi} \right.\right.\nonumber \\ 
&+& \left.\left.f^{aei}f^{dbi})\overline{\omega}^{aj}_{\mu}A^{e}_{\nu}\phi^{bj}_{\mu}\right]\left[D^{df}_{\nu}c^{f} + gf^{dfg}A^{f}_{\nu}c^{g} 
+ gf^{dfk}A^{k}_{\nu}c^{k}\right] + \left[2gf^{abk}\overline{\omega}_{\mu}^{ac}\partial_{\nu}\varphi^{bc}_{\mu} \right.\right.\nonumber \\ 
&+& \left.\left.2g^{2}(1-\eta)f^{dai}f^{dbk}\overline{\omega}^{ac}_{\mu}A^{i}_{\nu}\varphi^{bc}_{\mu} + g^{2}(1-\eta)f^{ade}f^{ebk}\overline{\omega}^{ac}_{\mu}A^{d}_{\nu}\varphi^{bc}_{\mu} 
+g\eta\overline{\omega}^{ac}_{\mu}f^{bak}\partial_{\nu}\varphi^{bc}_{\mu} \right.\right.\nonumber \\
&+& \left.\left.gf^{abk}\overline{\omega}^{ac}_{\mu}\phi^{bc}_{\mu}\partial_{\nu} + 2gf^{abk}\overline{\omega}^{aj}_{\mu}\partial_{\nu}\varphi^{bj}_{\mu} + 2g^{2}(1-\eta)f^{dai}f^{dbk}\overline{\omega}^{aj}_{\mu}A^{i}_{\nu}\varphi^{bj}_{\mu} \right.\right.\nonumber \\
&+& \left.\left.g^{2}(1-\eta)f^{ade}f^{cbk}\overline{\omega}^{aj}_{\mu}A^{d}_{\nu}\varphi^{bj}_{\mu} 
+ g\eta\overline{\omega}^{aj}_{\mu}f^{bak}\partial_{\nu}\varphi^{bc}_{\mu} + gf^{abk}\overline{\omega}^{aj}_{\mu}\varphi^{bj}_{\mu}\partial_{\nu} \right]\left[\partial_{\nu}c^{k} + gf^{fgk}A^{f}_{\nu}c^{g}\right] \right\}\;. \nonumber \\
\label{trividec2} 
\end{eqnarray}
A consistency check must be performed to verify if this trivial term interpolates among Landau and maximal Abelian gauges trivial terms \cite{Pereira:2013aza}. To avoid many tedious algebraic steps along the text, we leave this proof for the Ap.~\ref{trivialterms}.

\subsection{Breaking Term} \label{bterm}

We have now to introduce a soft BRST breaking term at the original action $S_0$, given by

\begin{equation}
S_0 = S_{\mathrm{YM}} + S_{\mathrm{LM}} + S_{triv}\;.
\label{initialaction}
\end{equation}
The reason is the following: Since we can obtain the Gribov copies equation requiring the BRST invariance of the gauge condition, to ruin this equation we must break the BRST invariance. This breaking, however, is not arbitrary. Therefore, when we look for perturbative effects of the theory, the BRST invariance must be restored, and for this reason, we call this a \textit{soft} breaking \cite{Baulieu:2008fy,Baulieu:2009xr}. A soft breaking can be obtained by the introduction of a mass parameter $\gamma$ which makes the dimension of this term lower than the spacetime dimension. See \cite{Pereira:2013aza} for more details about this construction. The general form of the soft breaking term is
\begin{eqnarray}
\Xi &=& \int d^{4}x \left[\gamma^{2}(D^{ab}_{\mu} + \xi(\eta) gf^{abc}A^{c}_{\mu})(\varphi + \overline{\varphi})^{ab}_{\mu} + \gamma^{2}\theta(\eta) g f^{aic}A^{c}_{\mu}(\varphi + \overline{\varphi})^{ai}_{\mu} \right.\nonumber \\
&+& \left.\gamma^{2}\chi(\eta)gf^{aic}A^{c}_{\mu}(\varphi + \overline{\varphi})^{ia}_{\mu} + \zeta_{1}(\eta)\gamma^{2}(\overline{\varphi}^{ab}_{\mu}\varphi^{ab}_{\mu} - \overline{\omega}^{ab}_{\mu}\omega^{ab}_{\mu}) + \zeta_{2}(\eta)\gamma^{2}(\overline{\varphi}^{ai}_{\mu}\varphi^{ai}_{\mu} - \overline{\omega}^{ai}_{\mu}\omega^{ai}_{\mu}) \right.\nonumber \\
&+& \left.\zeta_{3}(\eta)\gamma^{2}(\overline{\varphi}^{ib}_{\mu}\varphi^{ib}_{\mu} - \overline{\omega}^{ib}_{\mu}\omega^{ib}_{\mu}) + \zeta_{4}(\eta)\gamma^{2}(\overline{\varphi}^{ij}_{\mu}\varphi^{ij}_{\mu} - \overline{\omega}^{ij}_{\mu}\omega^{ij}_{\mu}) + \frac{\zeta_{5}}{2}(\eta)\gamma^{2}A^{a}_{\mu}A^{a}_{\mu} + \frac{\zeta_{6}}{2}(\eta)\gamma^{2}A^{i}_{\mu}A^{i}_{\mu} \right.\nonumber \\
&+& \left.\epsilon\gamma^{4}\right]\;,
\label{breakgeral}
\end{eqnarray}
where the parameters $\xi$, $\theta$, $\chi$ and $\zeta_{i}$ must be $\eta$-dependent in order to permit the interpolation of the breaking term of Landau and maximal Abelian gauges. Such interpolation is presented at Ap.~\ref{breaktermslM}. Comparing eq.(\ref{breakgeral}) with eq.(\ref{breaklandau}) and eq.(\ref{breakmag}), we obtain
\begin{eqnarray}
\xi(\eta) &=& \frac{1}{2}(1 - 3\eta)\;, \nonumber \\
\theta(\eta) &=& -\eta\;, \nonumber \\
\chi(\eta) &=& \eta\;, \nonumber \\
\zeta_{1}(\eta) &=& \zeta_{1}\;, \nonumber \\
\zeta_{2}(\eta) &=& \zeta_{1} \eta\;, \nonumber \\
\zeta_{3}(\eta) &=& \zeta_{1} \eta\;, \nonumber \\
\zeta_{4}(\eta) &=& \zeta_{1} \eta\;, \nonumber \\
\zeta_{5}(\eta) &=& \zeta_{2}\;, \nonumber \\
\zeta_{6}(\eta) &=& \zeta_{2} \eta,\;
\label{parameters}
\end{eqnarray}
with $\zeta_{1}$ and $\zeta_{2}$ being independent from $\eta$. We remark that, depending on the Ward identities, some of these parameters might be zero. Hence, the breaking term $\Xi$ with the appropriated fixed parameters is
\begin{eqnarray}
\Xi &=& \int d^{4}x \left[\gamma^{2}(D^{ab}_{\mu} + \frac{1}{2}(1-3\eta)gf^{abc}A^{c}_{\mu})(\varphi + \overline{\varphi})^{ab}_{\mu} - \gamma^{2}\eta g f^{aic}A^{c}_{\mu}(\varphi + \overline{\varphi})^{ai}_{\mu} \right.\nonumber \\
&+& \left.\gamma^{2}\eta gf^{aic}A^{c}_{\mu}(\varphi + \overline{\varphi})^{ia}_{\mu} + \zeta_{1}\gamma^{2}(\overline{\varphi}^{ab}_{\mu}\varphi^{ab}_{\mu} - \overline{\omega}^{ab}_{\mu}\omega^{ab}_{\mu}) + \zeta_{1}\eta\gamma^{2}(\overline{\varphi}^{ai}_{\mu}\varphi^{ai}_{\mu} - \overline{\omega}^{ai}_{\mu}\omega^{ai}_{\mu}) \right.\nonumber \\
&+& \left.\zeta_{1}\eta\gamma^{2}(\overline{\varphi}^{ib}_{\mu}\varphi^{ib}_{\mu} - \overline{\omega}^{ib}_{\mu}\omega^{ib}_{\mu}) + \zeta_{1}\eta\gamma^{2}(\overline{\varphi}^{ij}_{\mu}\varphi^{ij}_{\mu} - \overline{\omega}^{ij}_{\mu}\omega^{ij}_{\mu}) + \zeta_{2}\gamma^{2}A^{a}_{\mu}A^{a}_{\mu} + \zeta_{2}\eta\gamma^{2}A^{i}_{\mu}A^{i}_{\mu} \right.\nonumber \\
&+& \left.\epsilon\gamma^{4}\right]\;.
\label{breakfixed}
\end{eqnarray}
We remark that this term not only breaks the BRST in a soft manner, but also ensures that the copies equation is ruined. As discussed in Sect.~\ref{method}, this breaking term has some sort of ``freedom". In order to write an action free of infinitesimal copies, we could just introduce the following term
\begin{eqnarray}
\tilde{\Xi} &=& \int d^{4}x \left[\gamma^{2}(D^{ab}_{\mu} + \frac{1}{2}(1-3\eta)gf^{abc}A^{c}_{\mu})(\varphi + \overline{\varphi})^{ab}_{\mu} - \gamma^{2}\eta g f^{aic}A^{c}_{\mu}(\varphi + \overline{\varphi})^{ai}_{\mu} \right.\nonumber \\
&+& \left.\gamma^{2}\eta gf^{aic}A^{c}_{\mu}(\varphi + \overline{\varphi})^{ia}_{\mu}\right]\;.
\label{tildexilmaig}
\end{eqnarray}
The point here is that the terms given by $\Xi - \tilde{\Xi}$ permit a construction very close to the refined Gribov-Zwanziger action \cite{Dudal:2008sp}. In fact, to make contact between Eq.\eqref{breakfixed} and the LCO formalism, we should write the mass terms as independent masses $m_i=\zeta_{i}\gamma^{2}$ and deal with local composite operators and their condensation. An immediate difference between both methods relies on the gap equation, see Sect.~\ref{GAP}.

\section{Gluon Propagator} \label{propagatorsgluon}

It is remarkable that the elimination of infinitesimal Gribov copies, as an apparent technicality, brings rich effects to the physical properties of non-abelian gauge theories. Of course, they play a fundamental role for a consistent quantization, but their elimination provides a profound change in the gluon and ghost propagators, specially at the infrared regime. This is a well-known feature for Landau and maximal Abelian gauges, see for instance \cite{Gribov:1977wm,Capri:2006cz}. In fact, the inclusion of dimension 2 condensates makes the analytic result of the propagators to stay in harmony with lattice results \cite{Cucchieri:2007md,Cucchieri:2009zt,Bogolubsky:2007ud,Sternbeck:2007ug}. In this section, we compute the off-diagonal and diagonal gluon propagators for the interpolating gauge. As mentioned before, this provides a good way to test the free of copies action presented here, since this gauge could be implemented in the lattice. An interesting feature to study is the deformation of the propagators of the Landau gauge into the propagators of MAG, a property that could be investigated in the lattice.

The full action $S$ is given by
\begin{equation}
S = S_{\mathrm{YM}} + S_{\mathrm{LM}} + S_{\mathrm{ext}} + S_{triv} + \Xi\;.
\label{fullaction}
\end{equation}
For the gluon propagator, only the quadratic action $S_{\mathrm{q}}(A)$ is required\footnote{The auxiliary fields were integrated out from the path integral.},
\begin{eqnarray}
S_{\mathrm{q}} &=& \lim_{\alpha,\beta \to 0} \int d^{4}x \left[A^{a}_{\mu}\frac{1}{2}\left(\frac{\alpha - 1}{\alpha} \partial_{\mu}\partial_{\nu} - \delta_{\mu \nu}\partial^{2}\right) \delta^{ab}A^{b}_{\nu} + \frac{A^{i}_{\mu}}{(-\partial^{2} - \zeta_{1}\gamma^{2})}\gamma^{4}g^{2}N\delta^{ij}\partial^{2}\frac{A^{j}_{\mu}}{(-\partial^{2} - \zeta_{1}\gamma^{2})}\right.\nonumber \\
&+&  \left.A^{i}_{\mu}\frac{1}{2}\left(\frac{\beta - 1}{\beta}\partial_{\mu}\partial_{\nu} - \delta_{\mu \nu}\partial^{2}\right)\delta^{ij}A^{j}_{\nu} + \frac{A^{a}_{\mu}}{(-\partial^{2} - \zeta_{1}\gamma^{2})}\frac{\gamma^{4}g^{2}(1-3\eta)^{2}(N-2)\delta^{ab}\partial^{2}}{4}\frac{A^{b}_{\mu}}{(-\partial^{2} - \zeta_{1}\gamma^{2})} \right.\nonumber \\
&+& \left.\frac{A^{a}_{\mu}}{(-\partial^{2} - \zeta_{1}\gamma^{2}\eta)}2\gamma^{4}g^{2}\eta^{2}\delta^{ab}\partial^{2}\frac{A^{b}_{\mu}}{(-\partial^{2} - \zeta_{1}\gamma^{2}\eta)} + A^{i}_{\mu}\frac{2\gamma^{4}g^{2}N\delta^{ij}}{(-\partial^{2} - \zeta_{1}\gamma^{2})}A^{j}_{\mu} +  A^{a}_{\mu}\frac{4\gamma^{4}\eta^{2}g^{2}\delta^{ab}}{(-\partial^{2} - \zeta_{1}\eta\gamma^{2})}A^{b}_{\mu}\right.\nonumber \\
&+& \left.A^{a}_{\mu}\frac{\gamma^{4}g^{2}(1-3\eta)^{2}(N-2)\delta^{ab}}{2(-\partial^{2} - \zeta_{1}\gamma^{2})}A^{b}_{\mu}  +  \frac{A^{i}_{\mu}}{(-\partial^{2} - \zeta_{1}\gamma^{2})} \zeta_{1}\gamma^{6}g^{2}N\delta^{ij}\frac{A^{j}_{\mu}}{(-\partial^{2} - \zeta_{1}\gamma^{2})} + \frac{1}{2}A^{a}_{\mu}\gamma^{2}\zeta_{2}\delta^{ab}A^{b}_{\mu} \right.\nonumber \\
&+& \left.\frac{1}{2}A^{i}_{\mu}\gamma^{2}\zeta_{2}\eta\delta^{ij}A^{j}_{\mu} + \frac{A^{a}_{\mu}}{(-\partial^{2} - \zeta_{1}\gamma^{2})}\frac{\zeta_{1}\gamma^{6}g^{2}(1-3\eta)^{2}(N-2)\delta^{ab}}{4}\frac{A^{b}_{\mu}}{(-\partial^{2} - \zeta_{1}\gamma^{2})} \right.\nonumber \\
&+& \left.\frac{A^{a}_{\mu}}{(-\partial^{2} - \zeta_{1}\gamma^{2}\eta)}2\zeta_{1}\gamma^{6}\eta^{3}g^{2}\delta^{ab}\frac{A^{b}_{\mu}}{(-\partial^{2} - \zeta_{1}\gamma^{2}\eta)} \right]\;, \nonumber \\
\label{squad}
\end{eqnarray}
where $\alpha$ and $\beta$ are gauge parameters. The actual interpolating gauge is obtained at the limit $\alpha=\beta=0$. Taking the Fourier transform of Eq.~\eqref{squad}, we obtain the following expression,
\begin{eqnarray}
\Sigma_{\mathrm{q}} &=& \lim_{\alpha,\beta \to 0} \int \frac{d^{4}k}{(2\pi)^{4}} \left[A^{a}_{\mu}(k)\frac{1}{2}\left(\delta_{\mu \nu}k^{2} + \frac{1-\alpha}{\alpha} k_{\mu}k_{\nu}\right) \delta^{ab}A^{b}_{\nu}(-k) - A^{i}_{\mu}(k)\frac{\gamma^{4}g^{2}N\delta^{ij}k^{2}}{(k^{2} - \zeta_{1}\gamma^{2})^{2}}A^{j}_{\mu}(-k)\right.\nonumber \\
&+& \left.A^{i}_{\mu}(k)\frac{1}{2}\left(\delta_{\mu \nu}k^{2} + \frac{1-\beta}{\beta}k_{\mu}k_{\nu} \right)\delta^{ij}A^{j}_{\nu}(-k) - A^{a}_{\mu}(k)\frac{\gamma^{4}g^{2}}{4}\frac{(1-3\eta)^{2}(N-2)\delta^{ab}k^{2}}{(k^{2} - \zeta_{1}\gamma^{2})^{2}}A^{b}_{\mu}(-k) \right.\nonumber \\
&+& \left.\frac{1}{2}A^{i}_{\mu}(k)\gamma^{2}\zeta_{2}\eta\delta^{ij}A^{j}_{\mu}(-k) - A^{a}_{\mu}(k)2\frac{\gamma^{4}g^{2}\eta^{2}\delta^{ab}k^{2}}{(k^{2} - \zeta_{1}\gamma^{2}\eta)^{2}}A^{b}_{\mu}(-k) + A^{i}_{\mu}(k)\frac{2\gamma^{4}g^{2}N\delta^{ij}}{(k^{2} - \zeta_{1}\gamma^{2})}A^{j}_{\mu}(-k) \right.\nonumber \\
&+& \left.A^{a}_{\mu}(k)\frac{\gamma^{4}g^{2}(1-3\eta)^{2}(N-2)\delta^{ab}}{2(k^{2} - \zeta_{1}\gamma^{2})}A^{b}_{\mu}(-k)+A^{a}_{\mu}(k)\frac{4\gamma^{4}\eta^{2}g^{2}\delta^{ab}}{(k^{2} - \zeta_{1}\eta\gamma^{2})}A^{b}_{\mu}(-k)\right.\nonumber \\
&+& \left.A^{i}_{\mu}(k) \frac{\zeta_{1}\gamma^{6}g^{2}N\delta^{ij}}{(k^{2} - \zeta_{1}\gamma^{2})^{2}}A^{j}_{\mu}(-k) + \frac{1}{2}A^{a}_{\mu}(k)\gamma^{2}\zeta_{2}\delta^{ab}A^{b}_{\mu}(-k)  +  A^{a}_{\mu}(k)\frac{2\zeta_{1}\gamma^{6}\eta^{3}g^{2}\delta^{ab}}{(k^{2} - \zeta_{1}\gamma^{2}\eta)^{2}}A^{b}_{\mu}(-k)\right.\nonumber \\
&+&\left.A^{a}_{\mu}(k)\frac{\zeta_{1}\gamma^{6}g^{2}(1-3\eta)^{2}(N-2)\delta^{ab}}{4(k^{2} - \zeta_{1}\gamma^{2})^{2}}A^{b}_{\mu}(-k) \right]. 
\label{sigmaq}
\end{eqnarray}
It is not a difficult task to obtain the diagonal and off-diagonal gluon propagators from \eqref{sigmaq}. One has only to invert the corresponding wave operators in the usual way. The expressions for the diagonal and off-diagonal gluon propagators are, respectively,
\begin{eqnarray}
\langle A^{i}_{\mu}(k)A^{j}_{\nu}(-k)\rangle &=& \frac{(k^{2} - \zeta_{1}\gamma^{2})\delta^{ij}}{(k^{2}+\gamma^{2}\eta\zeta_{2})(k^{2}-\zeta_{1}\gamma^{2}) + 2\gamma^{4}g^{2}N}\left(\delta_{\mu \nu} - \frac{k_{\mu}k_{\nu}}{k^{2}}\right)\;,
\label{diagprop}\\
\langle A^{a}_{\mu}(k)A^{b}_{\nu}(-k)\rangle &=& \frac{2\delta^{ab}}{2(k^{2} + \gamma^{2}\zeta_{2}) + \frac{\gamma^{4}g^{2}(1-3\eta)^{2}(N-2)}{(k^{2}-\gamma^{2}\zeta_{1})} + \frac{8\gamma^{4}g^{2}\eta^2}{(k^{2}-\gamma^{2}\eta\zeta_{1})}}\left(\delta_{\mu \nu} - \frac{k_{\mu}k_{\nu}}{k^{2}}\right)\;,
\label{offdiagprop}
\end{eqnarray}
where the limits $\alpha=\beta=0$ were already taken. If we rename each mass term\footnote{In order to match the masses with the usual conventions, we must rename the terms involving $\zeta_1$ with a minus sign, \textit{i.e.} $-\gamma^2\zeta_{1} = m^2_1$ and the terms $\gamma^{2}\zeta_{2}$ are correctly identified with $m^2_2$.} which appears with a $\zeta_i$ parameter in terms of independent masses, we can write these propagators as
\begin{eqnarray}
\langle A^{i}_{\mu}(k)A^{j}_{\nu}(-k)\rangle &=& \frac{(k^{2} + m_1^{2})\delta^{ij}}{(k^{2}+m_2^{2}\eta)(k^{2}+m_1^2) + 2\gamma^{4}g^{2}N}\left(\delta_{\mu \nu} - \frac{k_{\mu}k_{\nu}}{k^{2}}\right)\;,
\label{diagprop01}\\
\langle A^{a}_{\mu}(k)A^{b}_{\nu}(-k)\rangle &=& \frac{2\delta^{ab}}{2(k^{2} + m_2^2) + \frac{\gamma^{4}g^{2}(1-3\eta)^{2}(N-2)}{(k^{2}+m_1^2)} + \frac{8\gamma^{4}g^{2}\eta^2}{(k^{2}+m_1^{2}\eta)}}\left(\delta_{\mu \nu} - \frac{k_{\mu}k_{\nu}}{k^{2}}\right)\;.
\label{offdiagprop01}
\end{eqnarray}

Since we are dealing with the interpolating gauge, we have to check the deformation of the propagators \eqref{diagprop} and \eqref{offdiagprop} among Landau and maximal Abelian gauges for $\eta = 1$ and $\eta = 0$, respectively. An important remark has to be done here: Explicit computations for the values of the mass parameters will give masses that implicitly depend on the gauge parameter $\eta$. This would allow to predict the masses in one gauge if we know the respective values in another\footnote{To compute the explicit value of the masses, the next step would be to renormalize the theory. This is a very tricky task due to the fact that the gauge fixing is non-linear. Just like the usual MAG, the non-linearity of this gauge requires the introduction of quartic ghost interactions which may be introduced accompanied by an extra gauge parameter (say, $\alpha^\prime$). The LMAIG is then embedded into a larger class of gauges characterized by two gauge parameters ($\eta$ and $\alpha^\prime$). See, for instance \cite{Capri:2008ak}. After renormalization, the original LMAIG is recovered by the limit $\alpha^\prime\rightarrow0$. This is a typical, but lengthy and intricate, method employed in non-linear gauges. Only by knowing the $\eta$-dependence of the masses, would be possible to predict the propagators in another gauge of the same family with the correct mass values. An example of the complexity of such approach can be found in \cite{Dudal:2003by}, where the condensate $A^2$ is computed at the linear covariant gauges (which is linear but carries a gauge parameter). In this example the mass value depends on the gauge parameter in a highly non-trivial way. For example, the explicit value of the condensate can actually be used to interpolate the condensate between the respective mass values at the Landau and Feynman gauges.}. However, for the present qualitative purposes, this dependence can be neglected.

Considering first the diagonal gluon propagator, it depends explicitly on $\eta$ only through a mass term, a fact that highlights the difference between the introduction of mass terms in the Landau and maximal Abelian gauges. For the former, we introduce a mass term for the off-diagonal and diagonal components of the gauge field, while for the later, we introduce a mass term just for the off-diagonal sector. This is explicit in Eq.~\eqref{breakfixed}. On the other hand, the off-diagonal sector changes in a much more non-trivial way. We will give some special attention to that. For $\eta = 0$, the propagator \eqref{offdiagprop} reduces to
\begin{equation}
\langle A^{a}_{\mu}(k)A^{b}_{\nu}(-k)\rangle = \frac{2\delta^{ab}(k^{2}-\gamma^{2}\zeta_{1})}{2(k^{2}+\gamma^{2}\zeta_{2})(k^{2}-\gamma^{2}\zeta_{1})+\gamma^{4}g^{2}(N-2)}\left(\delta_{\mu \nu} - \frac{k_{\mu}k_{\nu}}{k^{2}}\right)\;,
\label{propmag}
\end{equation}
which is exactly the off-diagonal gluon propagator for the maximal Abelian gauge \cite{Capri:2010an}. Choosing $\eta=1$, we have
\begin{equation}
\langle A^{a}_{\mu}(k)A^{b}_{\nu}(-k)\rangle = \frac{(k^{2}-\gamma^{2}\zeta_{1})\delta^{ab}}{(k^{2}-\gamma^{2}\zeta_{1})(k^{2}+\gamma^{2}\zeta_{2}) + 2\gamma^{4}g^{2}N}\left(\delta_{\mu \nu} - \frac{k_{\mu}k_{\nu}}{k^{2}}\right)\;.
\label{proplandau}
\end{equation}
which coincides with \eqref{diagprop}, as expected. Of course, the new features that all these calculations bring reside on different values for $\eta$ than $0$ and $1$. Hence, since $\eta\in\left[0,1\right]$, we can choose $\eta$ in such a way that an explicit continuous deformation of the off-diagonal propagator can be seen. Another remark is that for an arbitrary value of $\eta$ different from 0 and 1, we have the propagators for a gauge with non-hermitian Faddeev-Popov operator. This is a very important feature because, in usual approaches \cite{Gribov:1977wm,Zwanziger:1988jt,Dudal:2008sp}, the construction of these propagators would be very difficult (if not impossible). We also remark that expressions \eqref{diagprop01} and \eqref{offdiagprop01} are much more related to the usual results of the refined Gribov-Zwanziger approach where the mass parameters are independent of the Gribov parameter (at least in a tree-level analysis). As a final remark on the propagators: Since the gauge fixing of the non-Abelian sector is not the Landau gauge, it is expected a longitudinal contribution to the off-diagonal propagator. This contribution is expected to survive at the MAG limit, as predicted by lattice simulations \cite{Mendes:2006kc, Bornyakov:2003ee}. However, the tree-level propagators \eqref{offdiagprop01} and \eqref{propmag} is actually transverse. The reason is that the gauge fixing is non-linear, thus, the non-linear part of the gauge fixing can only be visualized through higher order contributions in the loop expansion. This is result is actually consistent with all tree-level analytical analysis of the Gribov problem in the MAG (See for instance \cite{Capri:2010an,Capri:2008ak}).

\section{A few words about the gap equation}\label{GAP}

As discussed in \cite{Pereira:2013aza}, the method applied in this work can provide a generalized gap equation. The reason why it ``can" lies on the fact that we have some freedom in the choice of the breaking term, as discussed in Sect.~\ref{method}. First, if we do not choose the mass terms which give rise to the refined Gribov-Zwanziger action, no modification to the gap equation emerges. We must remind that, if we opt for a minimal breaking of the BRST symmetry, these terms are excluded (at least the $A^2$ and $\overline{c}c$ which are not BRST invariant) and can be considered through the LCO formalism \cite{Dudal:2008sp,Dudal:2011gd}. On the other hand, we can include these terms for two reasons: (i) they are permitted by power counting and dimensional analysis and (ii) we recover the refined Gribov-Zwanziger propagators independently from the gap equation form. We could include them with an explicit dependence on $\gamma$ and, since the gap equation is obtained by minimizing the quantum action with respect to $\gamma^{2}$, all these terms will contribute to it, see \cite{Pereira:2013aza}. This is the difference to the usual gap equation which does not contain these terms. Alternatively,  these terms could also be included with independent mass parameters (\textit{i.e.}, with no explicit dependence on $\gamma$) and the gap equation would not be affected.

The main importance of these possibilities is that they can be very welcome, since the usual gap equation (at the Landau gauge, for instance) throws the theory right at the horizon, which is precisely the place where infinitesimal copies start to appear. The decision of how good is an alternative gap equation will rise with its physical effects and consistency checks. For obvious reasons, this analysis is left for future investigation.

\section{Conclusions}

In this work we applied the method developed in \cite{Pereira:2013aza} to eliminate infinitesimal Gribov copies from the interpolating Landau - maximal Abelian gauge. We obtained an action free of copies, given by Eq.~\eqref{fullaction}, which has the same structure of the refined Gribov-Zwanziger-type actions \cite{Dudal:2008sp,Dudal:2011gd}. After that, with a suitable choice of the interpolating parameter, we extracted the diagonal and off-diagonal gluon propagators and showed that the results reduce to the well-known propagators for the Landau and maximal Abelian gauge fixings.

Although the elimination of infinitesimal Gribov copies for the interpolating gauge is important and interesting by its own merit, it brings some new features for the general problem of dealing with Gribov copies. As discussed throughout Sect.~\ref{fpoperatorinter}, this gauge has a non-hermitian Faddeev-Popov operator, which means that no order relation can be established between its eigenvalues. Consequently, the possibility of constructing a well defined free of copies region in functional space is not evident. In this sense, the elimination performed here through the method developed in \cite{Pereira:2013aza} opens a new door to the understanding of non-hermitian Faddeev-Popov operators. Moreover, the propagators computed in Sect.~\ref{propagatorsgluon} are already in a form to be compared with a possible lattice simulation of the LMAIG.

There are many issues that should be addressed now. All of them deserves investigation. However, each of them is quite extend and intricate, and therefore are beyond the goals of this work. Nevertheless, they are left for future investigation. To cite a few interesting topics to be investigated, we can start with the renormalizability problem of action \eqref{fullaction}. As in the case of the MAG, many complications and extended expressions, due to extra quartic ghost interacting terms, are expected. Another important issue is the Abelian and non-Abelian ghost propagators, another task that should demand a laborious amount of computations (a smart attack would be to start with the $SU(2)$ case). A third problem to be studied is the comprehension of what could be, if any, an analogous Gribov region for this gauge and the interpolation between the known regions of Landau and maximal Abelian gauges. Finally, as discussed in Sect.~\ref{GAP}, the effect of a possible alternative gap equation has to be taken under consideration. This last question opens the possibility of introducing the refined mass parameters directly on the gap equation as a function of the Gribov parameter instead of a independent local composite operators condensation. Obviously, we would start this study at the Landau gauge, which is the gauge where the Gribov problem and its effects is most understood.

\appendix

\section{(non-)Hermiticity of $\nabla^{ab}$} \label{herm}

Let's define, using Eq.~\eqref{nablianaint}, an operator $\overline{\nabla}$ given by
\begin{equation}
\overline{\nabla}^{ab} \equiv \nabla^{ab} - \tilde{\nabla}^{ab}= - g^{2}\eta f^{acd}f^{dbi}A^{c}_{\mu}A^{i}_{\mu} - \eta gf^{aci}A^{i}_{\mu}D^{cb}_{\mu} - \eta g^{2}f^{adi}f^{cbi}A^{d}_{\mu}A^{c}_{\mu}\;.
\label{nablabarra}
\end{equation}
Since $\tilde{\nabla}$ is the MAG Faddeev-Popov operator, it is hermitian. Thus, if we prove that $\overline{\nabla}$ is hermitian, it is sufficient to say that $\nabla$ is also hermitian. To study this possibility, let us consider the following expression\footnote{All integrations in this section are performed in $d^4x$. We will omit this for simplicity.} 
\begin{equation}
\int \phi^{a}(\overline{\nabla}^{ab}\psi^{b})^\dagger = \int \phi^{a}(- \eta gf^{aci}A^{i}_{\mu}D^{cb}_{\mu}\psi^{b\dagger} - \eta g^{2}f^{adi}f^{cbi}A^{d}_{\mu}A^{c}_{\mu}\psi^{b\dagger}- g^{2}\eta f^{acd}f^{dbi}A^{i}_{\mu}A^{c}_{\mu}\psi^{b\dagger})\;.
\label{prova1}
\end{equation}
Now, we will consider the three terms of the \textit{rhs} of Eq.\eqref{prova1} separately. The first term is 
\begin{eqnarray}
\int \phi^{a}(- \eta gf^{aci}A^{i}_{\mu}D^{cb}_{\mu}\psi^{b\dagger}) &=& -\eta g\int \phi^{a}(f^{aci}A^{i}_{\mu}\delta^{cb}\partial_{\mu}\psi^{b\dagger} - gf^{aci}f^{cbj}A^{i}_{\mu}A^{j}_{\mu}\psi^{b\dagger}) \nonumber \\
&=& -\eta g\int f^{aci}\phi^{a}A^{i}_{\mu}\delta^{cb}\partial_{\mu}\psi^{b\dagger} + \eta g^{2}\int f^{aci}f^{cbj}\phi^{a}A^{i}_{\mu}A^{j}_{\mu}\psi^{b\dagger} \nonumber \\
&=& -\eta g\int f^{bci}\phi^{b}A^{i}_{\mu}\delta^{ca}\partial_{\mu}\psi^{a\dagger} + \eta g^{2}\int f^{bci}f^{caj}\phi^{b}A^{i}_{\mu}A^{j}_{\mu}\psi^{a\dagger} \nonumber \\
&=& \eta g\int f^{bci}(\partial_{\mu}\phi^{b})A^{i}_{\mu}\delta^{ca}\psi^{a\dagger} + \eta g^{2}\int \psi^{a\dagger} f^{cjb}f^{aci}\phi^{b}A^{i}_{\mu}A^{j}_{\mu}\phi^{b} \nonumber \\
&=& \int(-\eta g \psi^{a\dagger}f^{cai}A^{i}_{\mu}D^{cb}_{\mu}\phi^{b})\;, 
\label{prova2}
\end{eqnarray}
which proves its hermiticity. The second term is given by
\begin{eqnarray}
\int \phi^{a}(- \eta g^{2}f^{adi}f^{cbi}A^{d}_{\mu}A^{c}_{\mu}\psi^{b\dagger}) &=& \int \phi^{b}(- \eta g^{2}f^{bdi}f^{cai}A^{d}_{\mu}A^{c}_{\mu}\psi^{a\dagger}) \nonumber \\
&=& \int\phi^{b}(- \eta g^{2}f^{bci}f^{dai}A^{c}_{\mu}A^{d}_{\mu}\psi^{a\dagger}) \nonumber \\ 
&=& \int \psi^{a\dagger}(- \eta g^{2}f^{cbi}f^{adi}A^{d}_{\mu}A^{c}_{\mu}\phi^{b})\;, 
\label{prova3}
\end{eqnarray}
from where we can see that it is also hermitian. Finally, the third term is written as
\begin{eqnarray}
\int \phi^{a}(-\eta g^{2}f^{acd}f^{dbi}A^{i}_{\mu}A^{c}_{\mu}\psi^{b\dagger}) &=& \int \phi^{b}(-\eta g^{2}f^{bcd}f^{dai}A^{i}_{\mu}A^{c}_{\mu}\psi^{a\dagger}) \nonumber \\
&=& g^{2}\eta \int (f^{bad}f^{dic} + f^{bid}f^{dca})\phi^{b}\psi^{a\dagger}A^{i}_{\mu}A^{c}_{\mu} \nonumber \\
&=& g^{2}\eta \int f^{bad}f^{dic}\phi^{b}\psi^{a\dagger}A^{i}_{\mu}A^{c}_{\mu} - g^{2}\eta\int f^{dbi}f^{acd}\phi^{b}\psi^{a\dagger}A^{i}_{\mu}A^{c}_{\mu}\;. \nonumber \\
\label{prova4}
\end{eqnarray}
Clearly, the third term is not hermitian. Obviously, the reason is the presence of the piece
\begin{equation}
g^{2}\eta \int f^{bad}f^{dic}\phi^{b}\psi^{a\dagger}A^{i}_{\mu}A^{c}_{\mu}
\label{prova5}
\end{equation}
in Eq.~(\ref{prova4}). Since the sum of hermitian operators is a hermitian operator, we can see from expression \eqref{prova1} that $\overline{\nabla}$ would be hermitian only if \eqref{prova5} vanishes. The conclusion is that the purely off-diagonal components of the Faddeev-Popov operator of the Landau - MAG interpolating gauge is not hermitian, except for the MAG limit. In the case of the Landau gauge, this operator combines itself with the other sectors in order to provide another hermitian Faddeev-Popov operator.

\section{Trivial terms - Maximal Abelian Gauge} \label{trivialterms}

As discussed in Sect.~\ref{fpoperatorinter}, the Faddeev-Popov operator of the interpolating gauge has four different sectors: One purely off-diagonal, one purely diagonal and two mixed operators. The Landau gauge has this same feature and the comparison between the trivial terms of these gauges can be done term by term. Since we have seen that the operator given by Eq.~\eqref{fpinterpolante} reduces to the Faddeev-Popov operator at the Landau gauge for $\eta=1$, the expressions of the trivial term of the interpolating gauge, for $\eta = 1$, must also coincide with the trivial term of the Landau gauge (which must be decomposed for a comparison). This is easy to verify. On the other hand, the usual MAG \cite{Capri:2006cz,Capri:2010an} has only off-diagonal components for the Faddeev-Popov operator. In this sense, we have to be careful with the trivial term because, when we choose $\eta=0$ for the interpolating parameter, the mixed and purely diagonal components of the Faddeev-Popov operator will not vanish and this will provide a different trivial term with respect to the usual MAG case. Here, we have to remember that all mixed components of the Faddeev-Popov operator of the usual MAG are eliminated from the very beginning of the analysis of the Gribov problem (due to the redundant condition), see Sect.~\ref{MAG}. Hence, the trivial term of the MAG must have the form
\begin{equation}
S^{\mathrm{MAG}}_{triv} = - \int d^4x \left[ \overline{\varphi}^{ac}_{\mu}\nabla^{ab}\varphi^{bc}_{\mu} - \overline{\omega}^{ac}_{\mu}\nabla^{ab}\omega^{bc}_{\mu} - \overline{\omega}^{ac}_{\mu}(sA^{d}_{\nu})\frac{\delta \nabla^{ab}}{\delta A^{d}_{\nu}}\varphi^{bc}_{\mu} - \overline{\omega}^{ac}_{\mu}(sA^{i}_{\nu})\frac{\delta \nabla^{ab}}{\delta A^{i}_{\nu}}\varphi^{bc}_{\mu}\right]\;,
\label{strivmag}
\end{equation}
where $\nabla^{ab}$ is given by Eq.~\eqref{fpmag}. Now, if we consider the trivial term \eqref{trividec2} for the interpolating gauge and taking $\eta=0$, we have
\begin{eqnarray}
S_{triv} &=& - \int d^{4}x \left[\overline{\varphi}^{ac}_{\mu}\nabla^{ab}\varphi^{bc}_{\mu} +  \overline{\varphi}^{aj}_{\mu}\nabla^{ab}\varphi^{bj}_{\mu} + \overline{\varphi}^{ic}_{\mu}\nabla^{ib}\varphi^{bc}_{\mu} + \overline{\varphi}^{ic}_{\mu}\nabla^{ij}\varphi^{jc}_{\mu} + \overline{\varphi}^{ij}_{\mu}\nabla^{ib}\varphi^{bj}_{\mu} \right.\nonumber \\
&+& \left.\overline{\varphi}^{ij}_{\mu}\nabla^{ik}\varphi^{kj}_{\mu} - \overline{\omega}^{ac}_{\mu}\nabla^{ab}\omega^{bc}_{\mu} - \overline{\omega}^{aj}_{\mu}\nabla^{ab}\omega^{bj}_{\mu}  - \overline{\omega}^{ic}_{\mu}\nabla^{ib}\omega^{bc}_{\mu} - \overline{\omega}^{ic}_{\mu}\nabla^{ij}\omega^{jc}_{\mu} \right.\nonumber \\
&-& \left.\overline{\omega}^{ij}_{\mu}\nabla^{ib}\omega^{bj}_{\mu} - \overline{\omega}^{ij}_{\mu}\nabla^{ik}\omega^{kj}_{\mu} - \overline{\omega}^{ac}_{\mu}(sA^{d}_{\nu})\frac{\delta\nabla^{ab}}{\delta A^{d}_{\nu}}\varphi^{bc}_{\mu} - \overline{\omega}^{ac}_{\mu}(sA^{i}_{\nu})\frac{\delta\nabla^{ab}}{\delta A^{i}_{\nu}}\varphi^{bc}_{\mu} \right.\nonumber \\
&-& \left.\overline{\omega}^{aj}_{\mu}(sA^{d}_{\nu})\frac{\delta\nabla^{ab}}{\delta A^{d}_{\nu}}\varphi^{bj}_{\mu} - \overline{\omega}^{aj}_{\mu}(sA^{i}_{\nu})\frac{\delta\nabla^{ab}}{\delta A^{i}_{\nu}}\varphi^{bj}_{\mu} -  \overline{\omega}^{ic}_{\mu}(sA^{d}_{\nu})\frac{\delta\nabla^{ib}}{\delta A^{d}_{\nu}}\varphi^{bc}_{\mu} - \overline{\omega}^{ij}_{\mu}(sA^{d}_{\nu})\frac{\delta\nabla^{ib}}{\delta A^{d}_{\nu}}\varphi^{bj}_{\mu} \right]\;, \nonumber \\
\label{intmagtriv}
\end{eqnarray}
where the explicit expressions for $\nabla$ are
\begin{eqnarray}
\nabla^{ab} &=& -D^{ac}_{\mu}D^{cb}_{\mu} - gf^{acd}A^{c}_{\mu}D^{db}_{\mu} + g^{2}f^{adi}f^{cbi}A^{d}_{\mu}A^{c}_{\mu}\;, \nonumber \\
\nabla^{ai} &=& 0\;, \nonumber \\
\nabla^{ia} &=& gf^{abi}(\partial_{\mu}A^{b}_{\mu} + A^{b}_{\mu}\partial_{\mu})\;, \nonumber \\
\nabla^{ij} &=& - \delta^{ij} \partial^{2}\;.
\label{fpinterpolanteetazero}
\end{eqnarray}
Obviously, Eq.~\eqref{strivmag} is different from Eq.~\eqref{intmagtriv}, because the last term involves mixed and purely diagonal components of the auxiliary fields, which are absent at the usual MAG \eqref{strivmag}. However, it is possible to make the following change of variables in eq.(\ref{intmagtriv}): 
\begin{eqnarray}
\overline{\varphi}^{ac}_{\mu} &\longrightarrow& \overline{\varphi}^{ac}_{\mu} - \overline{\varphi}^{ic}_{\mu}\nabla^{id}(\nabla^{-1})^{da}\;,\nonumber  \\
\overline{\varphi}^{aj}_{\mu} &\longrightarrow& \overline{\varphi}^{aj}_{\mu} - \overline{\varphi}^{ij}_{\mu}\nabla^{id}(\nabla^{-1})^{da}\;,\nonumber  \\
\omega^{jc}_{\mu} &\longrightarrow& \omega^{jc}_{\mu} - (\nabla^{-1})^{jl}\nabla^{lb}\omega^{bc}_{\mu} - (\nabla^{-1})^{jl}(sA^{d}_{\nu})\frac{\delta\nabla^{lb}}{\delta A^{d}_{\nu}}\varphi^{bc}_{\mu}\;,\nonumber \\
\omega^{kj}_{\mu} &\longrightarrow& \omega^{kj}_{\mu} - (\nabla^{-1})^{kl}\nabla^{lb}\omega^{bj}_{\mu} - (\nabla^{-1})^{kl}(sA^{d}_{\nu})\frac{\delta\nabla^{lb}}{\delta A^{d}_{\nu}}\varphi^{bj}_{\mu}\;.\nonumber
\label{changevar}
\end{eqnarray}
The result is then
\begin{eqnarray}
S_{triv} &=& - \int d^{4}x \left[\overline{\varphi}^{ac}_{\mu}\nabla^{ab}\varphi^{bc}_{\mu} + \overline{\varphi}^{aj}_{\mu}\nabla^{ab}\varphi^{bj}_{\mu} + \overline{\varphi}^{ic}_{\mu}\nabla^{ij}\varphi^{jc}_{\mu} + \overline{\varphi}^{ij}_{\mu}\nabla^{ik}\varphi^{kj}_{\mu} - \overline{\omega}^{ac}_{\mu}\nabla^{ab}\omega^{bc}_{\mu} \right.\nonumber \\ &-& 
\left.\overline{\omega}^{aj}_{\mu}\nabla^{ab}\omega^{bj}_{\mu} -\overline{\omega}^{ic}_{\mu}\nabla^{ij}\omega^{jc}_{\mu} - \overline{\omega}^{ij}_{\mu}\nabla^{ik}\omega^{kj}_{\mu} - \overline{\omega}^{ac}_{\mu}(sA^{d}_{\nu})\frac{\delta\nabla^{ab}}{\delta A^{d}_{\nu}}\varphi^{bc}_{\mu} - \overline{\omega}^{ac}_{\mu}(sA^{i}_{\nu})\frac{\delta\nabla^{ab}}{\delta A^{i}_{\nu}}\varphi^{bc}_{\mu} \right.\nonumber \\ &-&\left.\overline{\omega}^{aj}_{\mu}(sA^{d}_{\nu})\frac{\delta\nabla^{ab}}{\delta A^{d}_{\nu}}\varphi^{bj}_{\mu} 
-\overline{\omega}^{aj}_{\mu}(sA^{i}_{\nu})\frac{\delta\nabla^{ab}}{\delta A^{i}_{\nu}}\varphi^{bj}_{\mu}\right]\;.
\label{trivchangedvar}
\end{eqnarray}
Evaluating the path integral to the mixed and purely diagonal fields by the use of the identity\footnote{Here, capital indices are used to remind that this identity holds for all combinations of diagonal and off-diagonal indices.} 
\begin{equation}
\int [\overline{\varphi}\varphi\overline{\omega}\omega]\exp\left\{-\int d^4x\left(\overline{\varphi}^{AC}_\mu\nabla^{AB}\varphi^{BC}_\mu-\overline{\omega}^{AC}_\mu\nabla^{AB}\omega^{BC}_\mu\right)\right\}=1\;,\label{idexp}
\end{equation} 
we finally obtain 
\begin{equation}
S_{triv} = - \int d^4x \left[ \overline{\varphi}^{ac}_{\mu}\nabla^{ab}\varphi^{bc}_{\mu} - \overline{\omega}^{ac}_{\mu}\nabla^{ab}\omega^{bc}_{\mu} - \overline{\omega}^{ac}_{\mu}(sA^{d}_{\nu})\frac{\delta \nabla^{ab}}{\delta A^{d}_{\nu}}\varphi^{bc}_{\mu} - \overline{\omega}^{ac}_{\mu}(sA^{i}_{\nu})\frac{\delta \nabla^{ab}}{\delta A^{i}_{\nu}}\varphi^{bc}_{\mu}\right]\;.
\label{strivmagint}
\end{equation}
Thus, we conclude that it is possible to achieve the known expression for MAG trivial term after a suitable change of variables and functional integration of the unwanted sectors.

\section{Breaking Terms - Landau and maximal Abelian gauges} \label{breaktermslM}

In Sect.~\ref{bterm}, we introduced the breaking term for the interpolating gauge. This term was introduced through the method described in Sect. \ref{method} and we fixed the interpolating parameters by comparing this term with the breaking terms of Landau and maximal Abelian gauges. In this appendix we provide the explicit expression of these terms. For the decomposed Landau ($\eta=1$) gauge, we have
\begin{eqnarray}
\Xi_{\mathrm{L}} &=& \int d^{4}x\left[\gamma^{2}(D^{ab}_{\mu} - gf^{abc}A^{c}_{\mu})(\varphi + \overline{\varphi})^{ab}_{\mu} - g\gamma^{2}f^{aic}A^{c}_{\mu}(\varphi + \overline{\varphi})^{ai}_{\mu} + g\gamma^{2}f^{aic}A^{c}_{\mu}(\varphi + \overline{\varphi})^{ia}_{\mu} \right.\nonumber \\
&+&\left.\zeta_{1}\gamma^{2}(\overline{\varphi}^{ab}_{\mu}\phi^{ab}_{\mu} - \overline{\omega}^{ab}_{\mu}\omega^{ab}_{\mu}) + \zeta_{1}\gamma^{2}(\overline{\varphi}^{ai}_{\mu}\varphi^{ai}_{\mu} - \overline{\omega}^{ai}_{\mu}\omega^{ai}_{\mu}) + \zeta_{1}\gamma^{2}(\overline{\varphi}^{ib}_{\mu}\varphi^{ib}_{\mu} - \overline{\omega}^{ib}_{\mu}\omega^{ib}_{\mu}) + \zeta_{1}\gamma^{2}(\overline{\varphi}^{ij}_{\mu}\varphi^{ij}_{\mu} \right.\nonumber \\
&-& \left.\overline{\omega}^{ij}_{\mu}\omega^{ij}_{\mu}) + \gamma^{2}\frac{\zeta_{2}}{2}A^{a}_{\mu}A^{a}_{\mu} + \gamma^{2}\frac{\zeta_{2}}{2}A^{i}_{\mu}A^{i}_{\mu} + \epsilon\gamma^{4}\right] \;,
\label{breaklandau}
\end{eqnarray}
and for the MAG ($\eta=0$),
\begin{equation}
\Xi_{\mathrm{MAG}} = \int d^{4}x \left[\gamma^{2}(D^{ab}_{\mu} + \frac{1}{2}gf^{abc}A^{c}_{\mu})(\overline{\varphi} + \varphi)^{ab}_{\mu} + \zeta_{1}\gamma^{2}(\overline{\varphi}^{ab}_{\mu}\varphi^{ab}_{\mu} - \overline{\omega}^{ab}_{\mu}\omega^{ab}_{\mu}) + \gamma^{2}\frac{\zeta_{2}}{2}A^{a}_{\mu}A^{a}_{\mu} + \epsilon\gamma^{4} \right]\;.
\label{breakmag}
\end{equation}
Both expressions are consistent with those obtained in \cite{Pereira:2013aza}.

\section*{Acknowledgements}

We acknowledge Anderson A. Tomaz and Bruno L. de Souza for their help with computational issues and Markus Huber for some comments on the manuscript. We are also very thankful to the referee for the fair criticisms and suggestions on the first version of this work. The Conselho Nacional de Desenvolvimento Cient\'{i}fico e Tecnol\'{o}gico\footnote{ RFS is a level PQ-2 researcher under the program \emph{Produtividade em Pesquisa}, 308845/2012-9.} (CNPq-Brazil) and the Pr\'o-Reitoria de Pesquisa, P\'os-Gradua\c c\~ao e Inova\c c\~ao (PROPPI-UFF) are acknowledge for financial support.

\end{document}